\documentclass[preprint2]{aastex}
\usepackage{natbib}
\usepackage{graphicx,graphics,subfigure}

\bibpunct{(}{)}{;}{a}{}{,}

\slugcomment{To appear in the Astronomical Journal}

\shorttitle{{\em Spitzer} observations of A1763: I}
\shortauthors{Edwards et al.}

\begin{document}

\title{{\em Spitzer} observations of Abell 1763 - \\
    I: infrared and optical photometry}

\author{Louise O. V. Edwards and Dario Fadda
\affil{NASA Herschel Science Center, Caltech 100-22, Pasadena, CA 91125}}
\email{louise@ipac.caltech.edu}

\author{Andrea Biviano}
\affil{INAF / Osservatorio Astronomico di Trieste, via G. B. Tiepolo 11, I-34143, Trieste, Italy}

\and

\author{Francine R. Marleau}
\affil{U.S. Planck Center, Caltech  220-6,  CA 91125 Pasadena}

\begin{abstract}
We present a photometric analysis of the galaxy cluster Abell 1763 at visible and infrared wavelengths. Included are fully reduced images in r$^{\prime}$, J, H, and K$_{s}$ obtained using the Palomar 200in telescope, as well as the IRAC and MIPS images from {\it Spitzer}. The cluster is covered out to approximately 3 virial radii with deep 24$\mu$m imaging (a 5$\sigma$ depth of 0.2 mJy). This same field of $\sim$40$^{\prime}$ by 40$^{\prime}$ is covered in all four IRAC bands as well as the longer wavelength MIPS bands (70 and 160$\mu$m). The r$^{\prime}$ imaging covers $\sim$0.8$\,$deg$^{2}$ down to 25.5 magnitudes, and overlaps with most of the MIPS field of view. The J, H, K$_{s}$ images cover the cluster core and roughly half of the filament galaxies, which extend towards the neighboring cluster, Abell~1770. This first, in a series of papers on Abell~1763, discusses the data reduction methods and source extraction techniques used for each dataset. We present catalogs of infrared (IR) sources (with 24 and/or 70$\mu$m emission) and their corresponding emission in the optical (u$^{\prime}$, g$^{\prime}$, r$^{\prime}$, i$^{\prime}$, z$^{\prime}$), and Near- to Far-IR (J, H, K$_{s}$, IRAC, and MIPS 160$\mu$m). We provide the catalogs and reduced images to the community through the NASA/IPAC Infrared Science Archive ({\em IRSA}).

\end{abstract}


\section{Introduction}

Whereas the local cluster environment is known to be harsh and generally unfavorable to star-forming processes \citep{abr96,van98,bal99}, the observational evidence for increased activity in lower density  environments is plentiful, at least in the local universe \citep{dre80,dre97,gom03}. As one looks across the diameter of a
galaxy cluster, more often the star-forming galaxies are found in the field rather than at the cluster center, except particular
cases such as interacting galaxies and BCGs at the centers of cool cores \citep[and references therein]{edw07}. Recent studies emphasize that this higher fraction of star-forming galaxies occurs at intermediate galaxy density regions too, such as in groups and along filaments \citep{bal04,gal09}.

In dusty galaxies, the ultra-violet light emitted by young stars is reprocessed to longer wavelengths, and the infrared (IR) emission provides a direct window onto activity heavily obscured in the optical. Our {\em Spitzer} observations of the galaxy cluster Abell 1763 were the first to directly detect galaxies along a cluster filament using 24$\mu$m observations \citep{fad08}. With $>$1000 optical spectra, and 24$\mu$m images from the {\em Spitzer} Space Telescope, this letter reported the discovery of galaxies extending out to $\sim$12Mpc from Abell~1763 to the neighboring cluster Abell~1770. More importantly, the fraction of star-forming galaxies as a function of location within the superstructure was measured. It was found that the fraction of star-forming galaxies is twice as high in the filaments than in any other region.  \citet{gal09} has found increased yet obscured star-formation activity among galaxies in intermediate density environments in the A901/A902 supercluster, presumably caused by galaxy-galaxy interactions. Together with optical observations \citep{bra07,por07,por08} these studies show an enhanced fraction of star-forming galaxies, and an increased specific star-formation rate, among galaxies in the intermediate-density environments surrounding clusters. 

The discoveries outlined above support the results of cosmological simulations which build clusters through the streaming of galaxies into 
higher density environments via filaments \citep{van93}. Although the lower activity of cluster galaxies may simply be a manifestation of one form of downsizing \citep{tre05}, several mechanisms by which galaxies can, at least temporarily,
undergo an enhancement of star-formation are plausible in low and intermediate density environments. For example, gas in infalling galaxies can be
compressed via ram pressure with the intracluster medium, before it is stripped from
the galaxy \citep{gun72,gav85}. Also, gravitational interactions between filament
galaxies should be frequent as the velocity dispersion of the galaxies in pairs and groups
is low enough that mergers (or harassment) can upset the
stability of the galaxy's gas. Gas may then funnel towards the center of
 the galaxy triggering a star-formation episode \citep{mih94,mor05}. For this reason that it is important to study wide fields out past the cluster virial radius. Such a study allows examination of the outskirts and large scale structure, as well as comparison to the higher galaxy density environment of the cluster core.

With this in mind, our approach is to study individual clusters in detail,
as they are
insensitive to any cosmological evolution (such as a sample taken
within several clusters of different masses or redshifts) and present a large number of galaxies in a wide range in galaxy
density along the galaxy filaments (typically 2-8
$\rho_C$, where $\rho_C$ is the critical density of the Universe,
\citet{col05} to the cluster cores ($>$200 $\rho_C$)). Abell 1763,
a rich galaxy cluster at a redshift of z$\sim$0.2 makes up part of the superstructure which includes a neighboring poor cluster, Abell 1770 at a similar redshift, and the galaxy filament which connects the two. 

\nocite{ade08}
This is the first in a series of papers which present an in depth study of Abell 1763. As announced in the letter from \citet{fad08}, this is a local cluster in which filament galaxies are clearly more active than galaxies in any other region of the cluster. In this first paper, we present two photometric catalogs. The first is that of 24$\mu$m infrared-selected sources and their optical magnitudes. We obtain the optical associations from our own deep r$^{\prime}$ imaging as well as from the 7th Data Release of the Sloan Digital Sky Survey (Adelman-McCarthy et al. 2008; hereafter, SDSS), the Near-IR (JHK$_{s}$) magnitudes we obtain from the Palomar 200in telescope, and the Mid- to Far-IR fluxes (IRAC 3.6, 4.5, 5.8, 8.0$\mu$, and MIPS 24, 70 and 160$\mu$m) are from {\em Spitzer}. The second is a catalog of 70$\mu$m sources that are outside of the field of view of the 24$\mu$m image. This second catalog extends the field of view by about $\sim$30\% of what was considered in the letter. Both catalogs are based on {\em Spitzer} images which have been completely re-reduced. We describe our improved data reduction methods in detail for the IRAC, MIPS 24$\mu$m, and MIPS 70$\mu$m fields, as well as the reduction of the MIPS 160$\mu$m, whose analysis was also not included in the letter. The J, H, and K$_{s}$ data is also new since the letter. Future articles will describe the
spectroscopic follow-up observations of the infrared sources, and the estimates of star-formation rates and
stellar masses. At the time of this publication, we are currently examining UV (GALEX) and radio (VLA) data which we will use in the future to further investigate the total star-formation rates.

\section{{\em Spitzer} observations}

\begin{figure*}
\epsscale{1.8}
\plotone{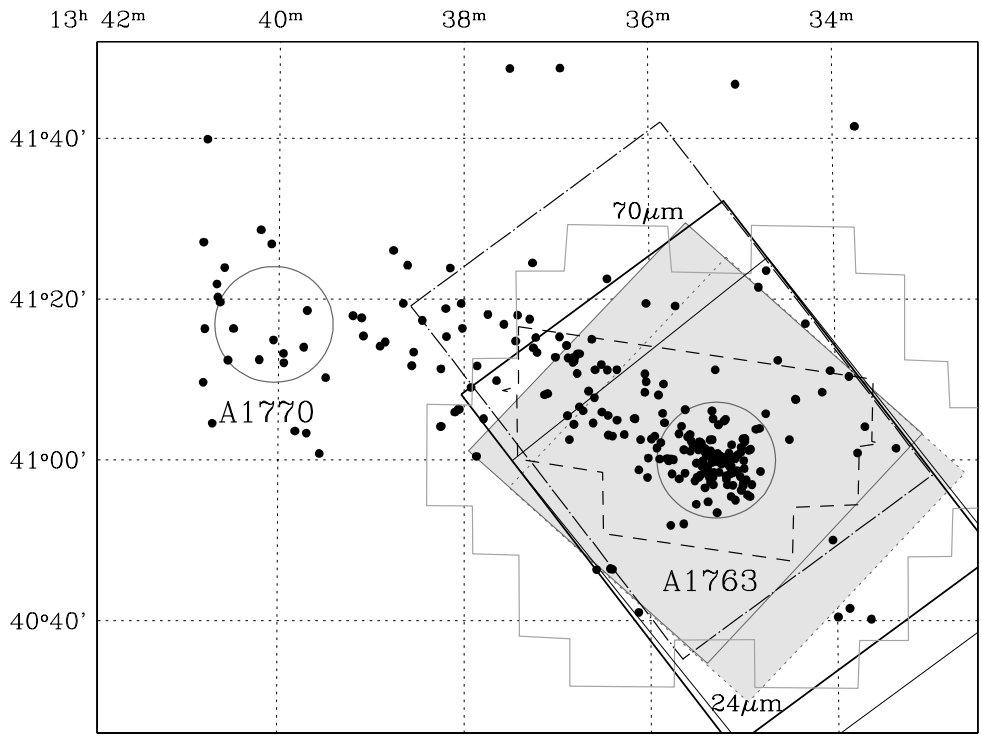}
\caption{{\bf Coverage of IR and optical data.}  Filled dots indicate spectroscopically confirmed members of the A1763-A1770 supercluster. The circles centered on the clusters have a radius of 1.5Mpc. The solid grey line marks the position of the LFC r$^{\prime}$ mosaic, the dashed polygon marks the WIRC mosaic. The shaded diamonds show the IRAC fields. The 3.6 and 5.8$\mu$m data are outlined with a dotted line and the 4.5 and 8.0$\mu$m data are outlined with a grey solid line. The large thick solid rectangle marks the 24$\mu$m observations, the large dot-dashed rectangle marks the 70$\mu$m data, and the large thin solid rectangle marks the 160$\mu$m observations. 
\label{spitzer_cov}
}
\end{figure*}

We imaged the Abell 1763 cluster in all the IRAC and MIPS
channels.  The different observations are summarized in
Table~\ref{obs} and the coverage of the 
observations at each wavelength is displayed in Figure~\ref{spitzer_cov}. For the {\it Spitzer} fluxes, the depth measurements are calculated from the average value of the RMS image (calculated as described in Section~\ref{Images}) and scaled by five. For the WIRC and LFC depths, we use the magnitude at which the number counts begin to drop. The FWHM for the ${\it Spitzer}$ observations is quoted from the Spitzer Data Handbook, version 8.0~\footnote{see p. 72 available http://ssc.spitzer.caltech.edu/mips/dh/}, and measured across bright stars in the frames for the WIRC and LFC observations. 

As we do not in general use the available pipeline final data products or processing software, we describe our reduction technique in detail in the following section. We also describe the methods used for  source extraction and photometry.

\begin{deluxetable}{lccccccc}
\tabletypesize{\scriptsize}
\tablewidth{0pt}
\tablecaption{Observations \label{obs}}
\tablehead{\colhead{INST} & \colhead{$\lambda$$_{cent}$ ($\mu$m)} & \colhead{ID} & \colhead{Date} & \colhead{Time(min)}& \colhead{Coverage($^{\prime 2}$)} & \colhead{Depth} &\colhead{FWHM} ($^{\prime\prime}$)}
\startdata
IRAC & 3.6 &r14790144 &   2005 Jun 13 & 76.8 & 1600&2.2$\mu$Jy 5$\sigma$&1.66\\
IRAC & 4.5 &r14790144  & 2005 Jun 13&76.8& 1600&4.6$\mu$Jy 5$\sigma$&1.72\\
IRAC & 5.8 &r14790144  & 2005 Jun 13&76.8& 1600&4.5$\mu$Jy 5$\sigma$&1.88\\
IRAC & 8.0 &r14790144  & 2005 Jun 13&76.8& 1600&4.3$\mu$Jy 5$\sigma$&1.98\\
MIPS & 24 &r14790912&   2005 Jun 28 & 171.0 & 2200 &0.2 mJy 5$\sigma$&5.9\\
MIPS & 24 &r14790912&   2005 Jun 28 & 171.0 & 2200 &0.2 mJy 5$\sigma$&5.9\\
MIPS & 24 &r14790912&   2005 Jun 28 & 171.0 & 2200 &0.2 mJy 5$\sigma$&5.9\\
MIPS     & 70 &14791168  &  2005 Jun 25 & 171.0& 2200&6.1 mJy 5$\sigma$&16\\
MIPS     & 70 &14791168  &  2005 Jun 25 & 171.0& 2200&6.1 mJy 5$\sigma$&16\\
MIPS     & 70 &14791168  &  2005 Jun 25 & 171.0& 2200&6.1 mJy 5$\sigma$&16\\
MIPS     & 160 &r14791424  &  2005 Jun 25 & 171.0& 2000&85.1 mJy 5$\sigma$&40\\
MIPS     & 160 &r14791424  &  2005 Jun 25 & 171.0& 2000&85.1 mJy 5$\sigma$&40\\
MIPS     & 160 & r14791424  &  2005 Jun 25 & 171.0& 2000&85.1 mJy 5$\sigma$&40\\
WIRCJ & 1.250 &     ...     &  2007 Mar 26 & 78.0& 1330& 19.5 mag$_{Vega}$&1.3\\
WIRCH & 1.635  &    ...     &  2007 Mar 26 & 39.3& 1330& 19.3 mag$_{Vega}$&1.2\\
WIRCK$_{s}$ & 2.150   &    ...    &  2007 Mar 26 & 22.2& 1330& 19.0 mag$_{Vega}$&1.4\\
LFC r$^{\prime}$ & 0.6255 &    ...      &   2004 Apr 20 &  2.5& 2880& 25.5 mag$_{AB}$&1.3\\
\enddata
\end{deluxetable}

\subsection{IRAC Data Reduction}

Basic calibrated datasets (BCDs) were downloaded from the {\em Spitzer}
archive. The BCDs used are the files for which basic data reduction steps, such
as dark subtraction, muxbleed, detector linearization, flat
fielding, and cosmic ray detection have been completed from within the
{\em Spitzer} pipeline (version S14.0.0). We further treat the BCDs with our
own artifact correction codes, significantly reducing several systematic and
cosmetic effects. For the four IRAC bands this includes corrections
for the column pulldown, jailbars, zodiacal light, subtraction of bright
stars and of the background level, as well as a treatment for the background droop near bright stars and glitches caused by cosmic ray hits. Before extracting the sources, stray light is removed from each of the BCDs which are then mosaiced together. The only alteration to this sequence is that
for IRAC 5.8 and 8.0$\mu$m  we include a correction for the superflat to remove the gradient inside each BCD before subtracting the bright stars.

\subsubsection{The column pulldown}

\begin{figure*}
 \subfigure{
 \begin{minipage}[c]{0.3\textwidth}
        \centering
        \label{1060conti}
        \includegraphics[width=2in,angle=0]{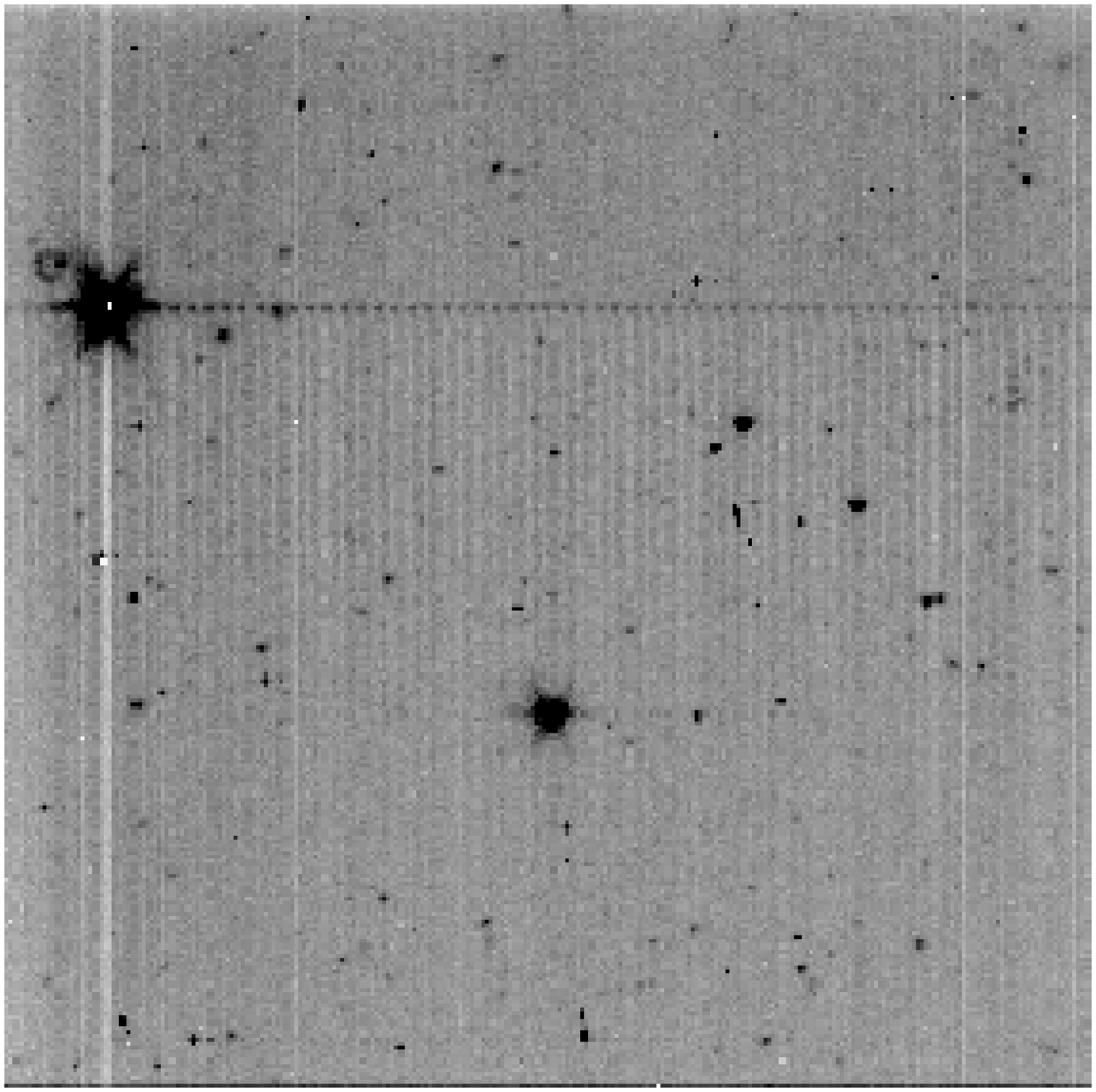}
        \end{minipage}%
    }
    \subfigure{
     \begin{minipage}[c]{0.3\textwidth}
        \centering
        \label{1060hai}
        \includegraphics[width=2in,angle=0]{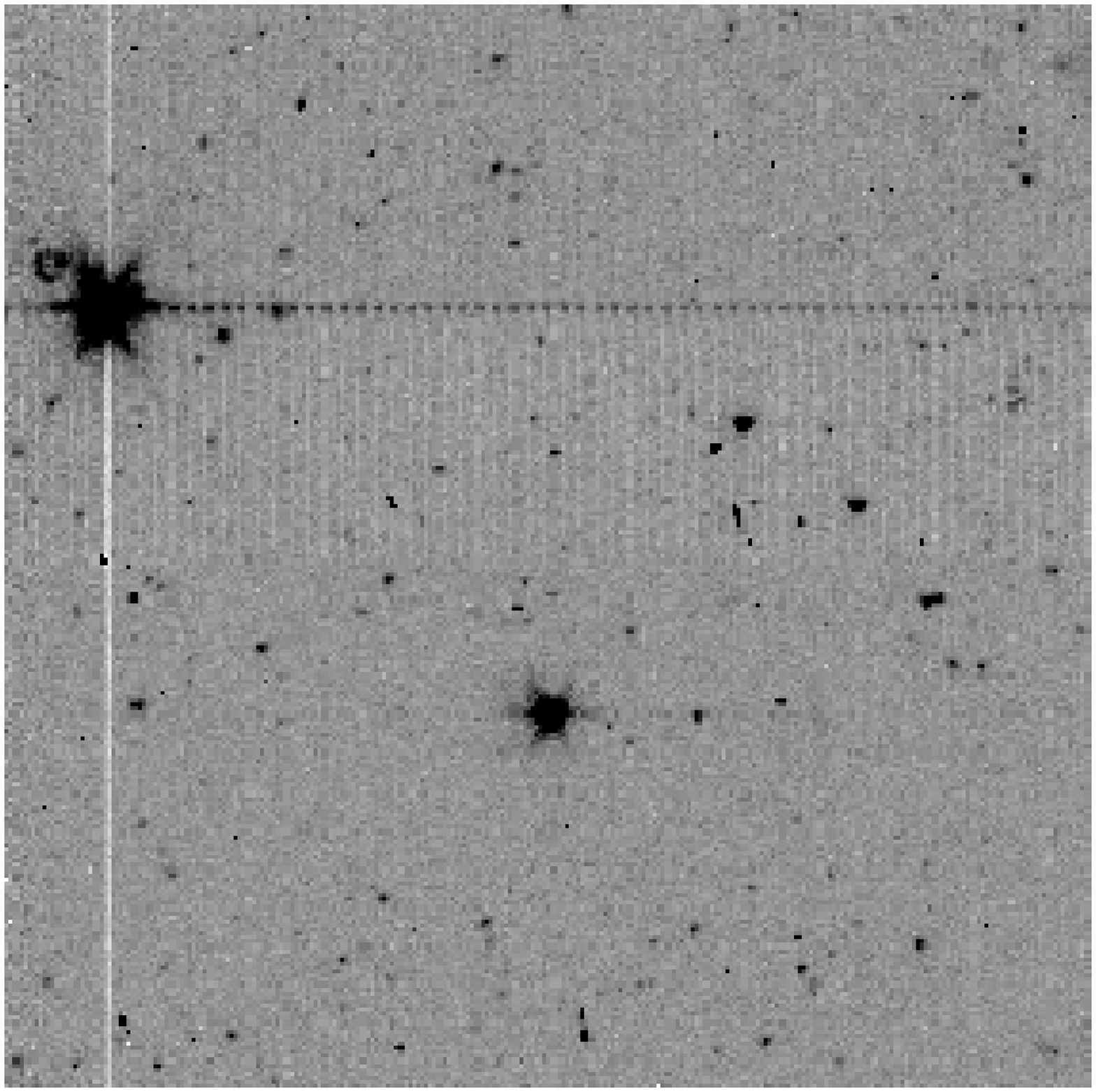}
         \end{minipage}%
    }
 \subfigure{
 \begin{minipage}[c]{0.3\textwidth}
        \centering
        \label{n2}
        \includegraphics[width=2in,angle=0]{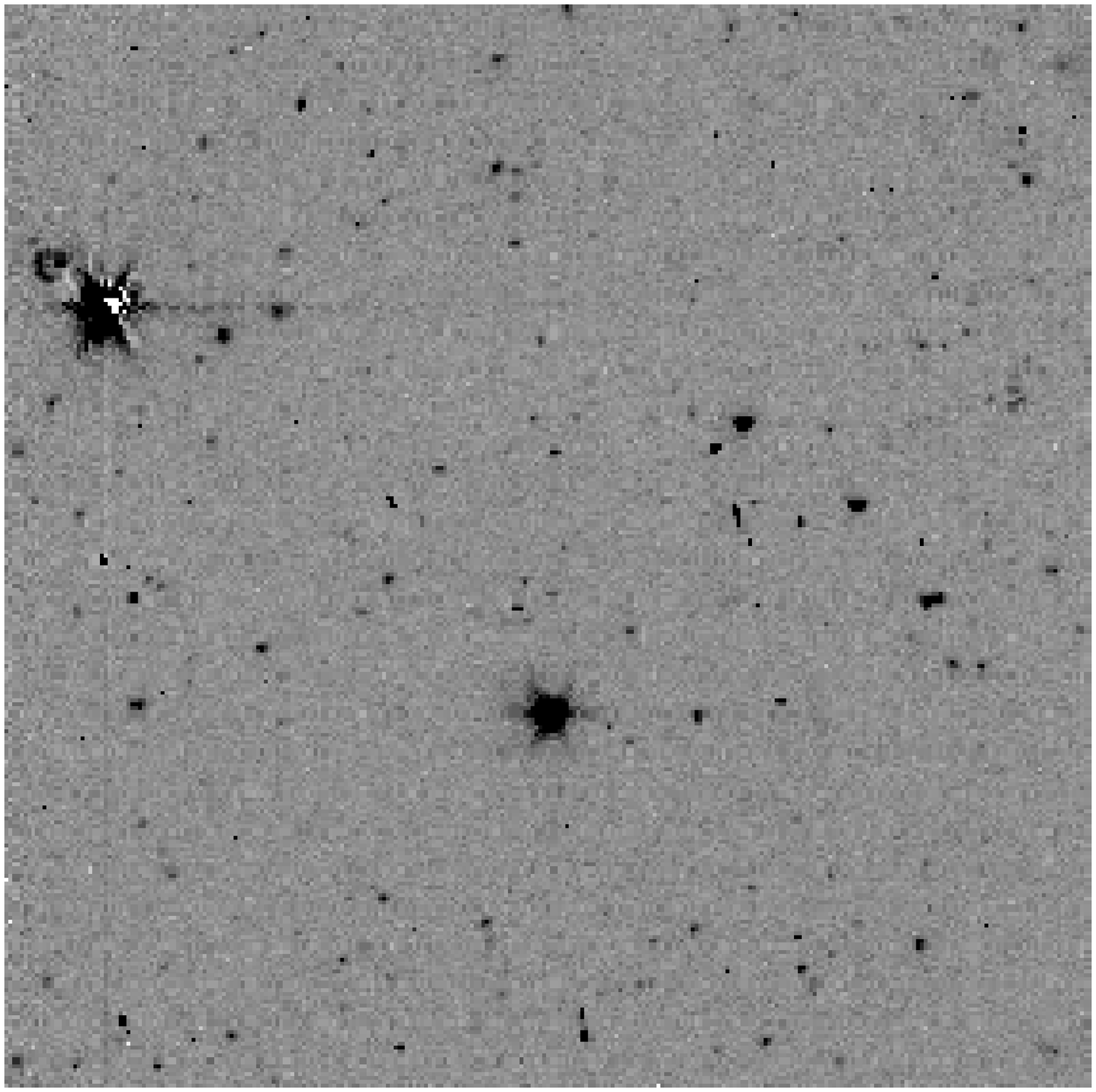}
        \end{minipage}%
    }
    \caption[]{{\bf BCD 58 from IRAC 3.6$\mu$m.} This frame illustrates the
    benefit of including a correction for the Jailbar (repeating bright and dark columns) and Column Pulldown (very low column values at the position of the bright star on the left hand side of the image)
    defects, for a single BCD, 5.2$^{\prime}$ on a side. The image includes a bright star where the effect is particularly apparent. From Left to Right: The raw image, the BCD after the
    pipeline corrections, the BCD after our additional corrections
    targeting the Jailbar and Column Pulldown defects.} \label{bcd58}
\end{figure*}

The column pulldown is an artifact that originates from very bright sources (stars, 
galaxies and cosmic ray hits) and causes a dip in the flux in the one or two columns
the bright source occupies. It is well illustrated in the first two
panels of Figure~\ref{bcd58}.  Our column pulldown correction works by first
masking sources in the field that are brighter than the biweight mean \citep{bee90} value by 5$\sigma$. We then recalculate the biweight mean for each column resulting in a set of 255 mean values for each BCD. The result is passed through a wavelength filter which excludes points that lie outside 8$\sigma$ of the mean. This disregards the small column to column variations and keeps only the columns where huge dips in flux occur (the columns which have
been pulled down by at least 8$\sigma$). We identify the column with the lowest average value and add to its pixel values, as well as those in the surrounding 2 columns, the difference between it and the local average of the bright-source-masked image. If there is more than one
pulldown column, and if they are adjacent to each other, we apply the
correction using the pixels bordering the outermost pulldown
column. The portions of the column
above the bright source and below the bright source have different
column pulldown intensities, therefore we do the correct separately for
these two regions. The maximum value of the pulldown column which separates the two regions, is taken to be the position of the star. As we identify the pulldown by the effect of the column, we will by definition catch all pulldown columns that are greater than 8$\sigma$ of the BCD background. It is possible smaller pulldown effects are missed. However, the pulldown is known to occur only from very bright stars or cosmic rays ($>$35000DN)~\footnote{Available http://ssc.spitzer.caltech.edu/irac/calib/features.html}. The final panel of Figure~\ref{bcd58} includes our
column pulldown correction. It is a particularly bad case with multiple
pulldown columns beside each other. Although the final panel does
indeed show some residual effect, the result is a major improvement compared to 
the BCD pipeline.

\subsubsection{The jailbars}

The second obvious and widespread cosmetic defect appearing in
these images is the jailbars effect, also appearing near bright
sources. The effect is a repeat of excessively bright and dark columns occurring
every 4th column.  It most noticeably manifests itself in the
row of the bright source as bright dots emanating from the
source. The first panel of Figure~\ref{bcd58} shows the pattern across the entirety of the BCD. It is particularly important to remove this effect
before building source catalogs, as these dots can easily be
mistaken for real sources by the automatic source detection
algorithms.

The jailbar correction is performed on to each BCD individually, as was the case for the pulldown correction. This effect is not as localized as the above column pulldown effect, so it
becomes important to disregard the diffuse light from large galaxies that
permeates throughout many of the BCDs. We start by masking the strong
sources and setting their value to the median value of the
image, we also give the median value to the image edges. What remains is the diffuse light with the strong sources masked to the median value. We subtract this from the original BCD leaving only the bright sources - the diffuse
light is subtracted. The jailbar effect occurs every 4th column, so the BCD is split into 4 sets of columns. For each of these, a median column profile is computed and then filtered through a wavelength algorithm.  The smoothed profiles containing the jailbar pattern are then subtracted from the 4 sets of columns. Finally, we add back the median value and the diffuse light.  The second panel of Figure~\ref{bcd58} shows how the BCD pipeline correction does a good job in the bottom half of the frame, and how our method removes most of the jailbars.

\subsubsection{The background and the effects of bright sources}

Our field of view is large enough that several bright stars are found
in the images. These bright stars do more than just enhance the
jailbar and column pulldown effects. They also cause a distortion in the
background and raise the overall average background level of the image.

  \begin{figure}[h]
  \epsscale{0.9}
     \plotone{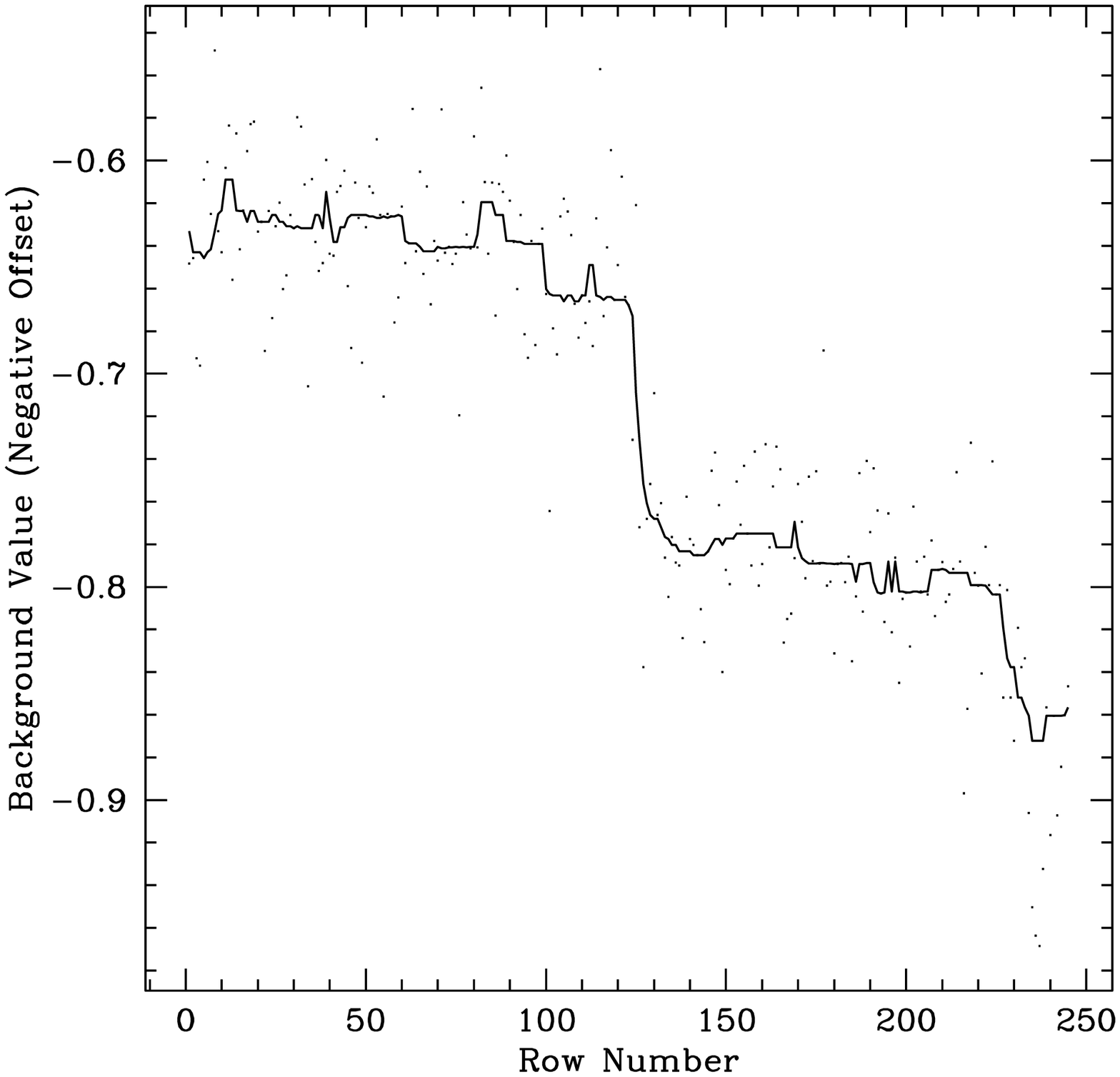}
    \caption{{\bf  The droop effect in IRAC 3.6$\mu$m.} The variation in flux along rows of pixels in frames
    with bright stars shows a droop across the frame. The points represent the median value of each row across one such frame. The star is located at rows 125-130. The line shows the profile that has been subtracted from the frame, which is the filtered profile including only points within 5$\sigma$. The flux unit is arbitrary and shows a negative offset as this step occurs before the matching the backgrounds of all the frames. \label{drooppl}}
\end{figure}

To remove the stars themselves, we download the point response
function (PRF) files that are available on the {\em Spitzer} homepage, one
for each IRAC band. The model PRFs have a resolution 5 times that of the BCDs so first we degrade the quality of the PRFs. There are four bright stars (with
V$<$11.3) that cause distortion. We subtract the degraded model PRFs from the position of the bright stars and re-run the column pulldown at the position of the stars on the
corrected image, as well as a column pullup, to account for the bright
columns around the stars.

 We remove the background of each BCD. First bright sources ($>$5$\sigma$) and the diffuse light are isolated and given a median value. We then median-filter (width=10) the BCD. We take the background value to be the minimum of the ``smoothed'' BCD, and subtract this value from the unmasked original.

\begin{figure*}
 \subfigure{
     \begin{minipage}[l]{0.45\textwidth}
        \centering
        \label{1060hai}
        \includegraphics[height=3.5in,angle=0]{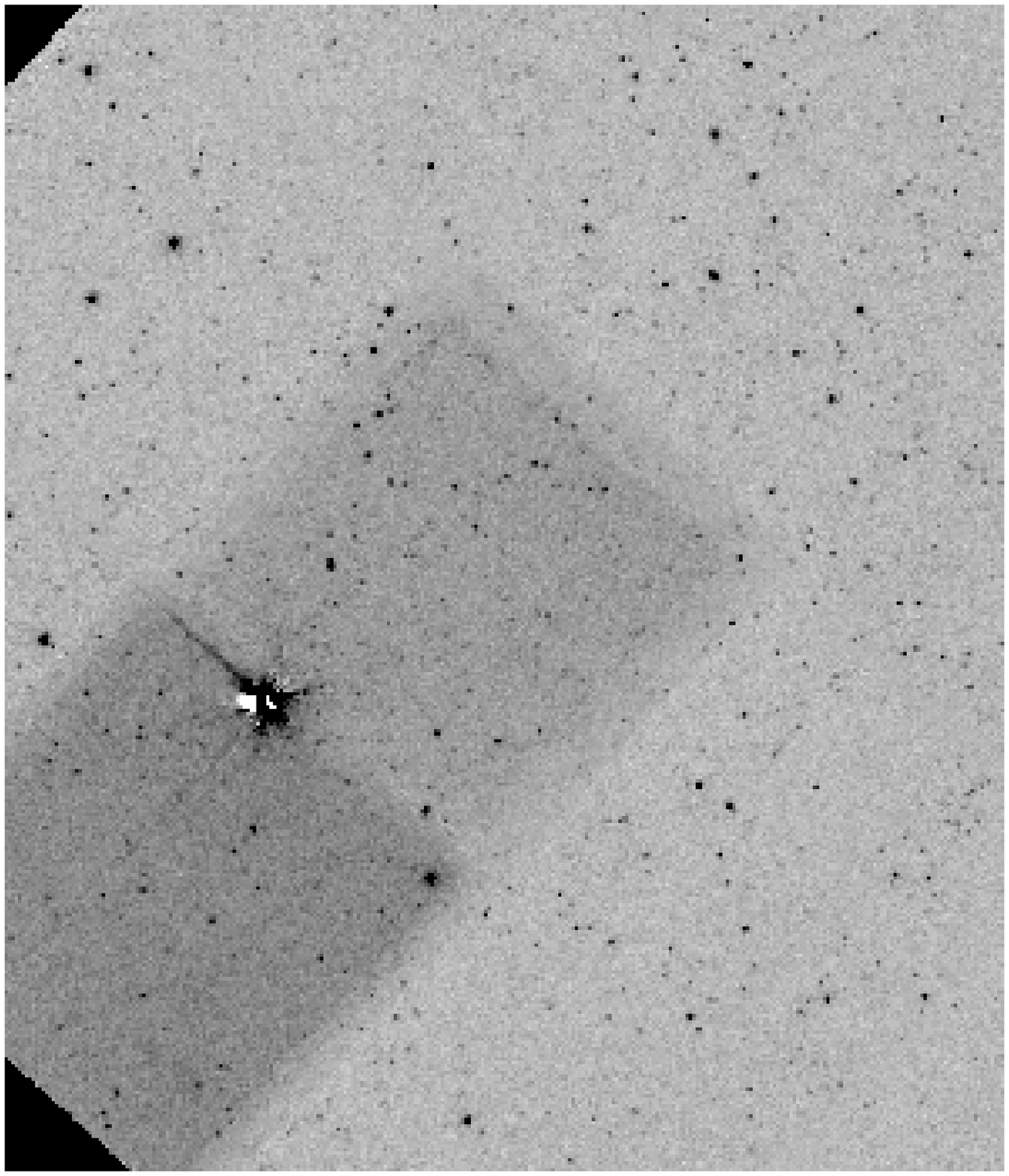}
         \end{minipage}%
    }
 \subfigure{
 \begin{minipage}[l]{0.55\textwidth}
        \centering
        \label{n2}
        \includegraphics[height=3.5in,angle=0]{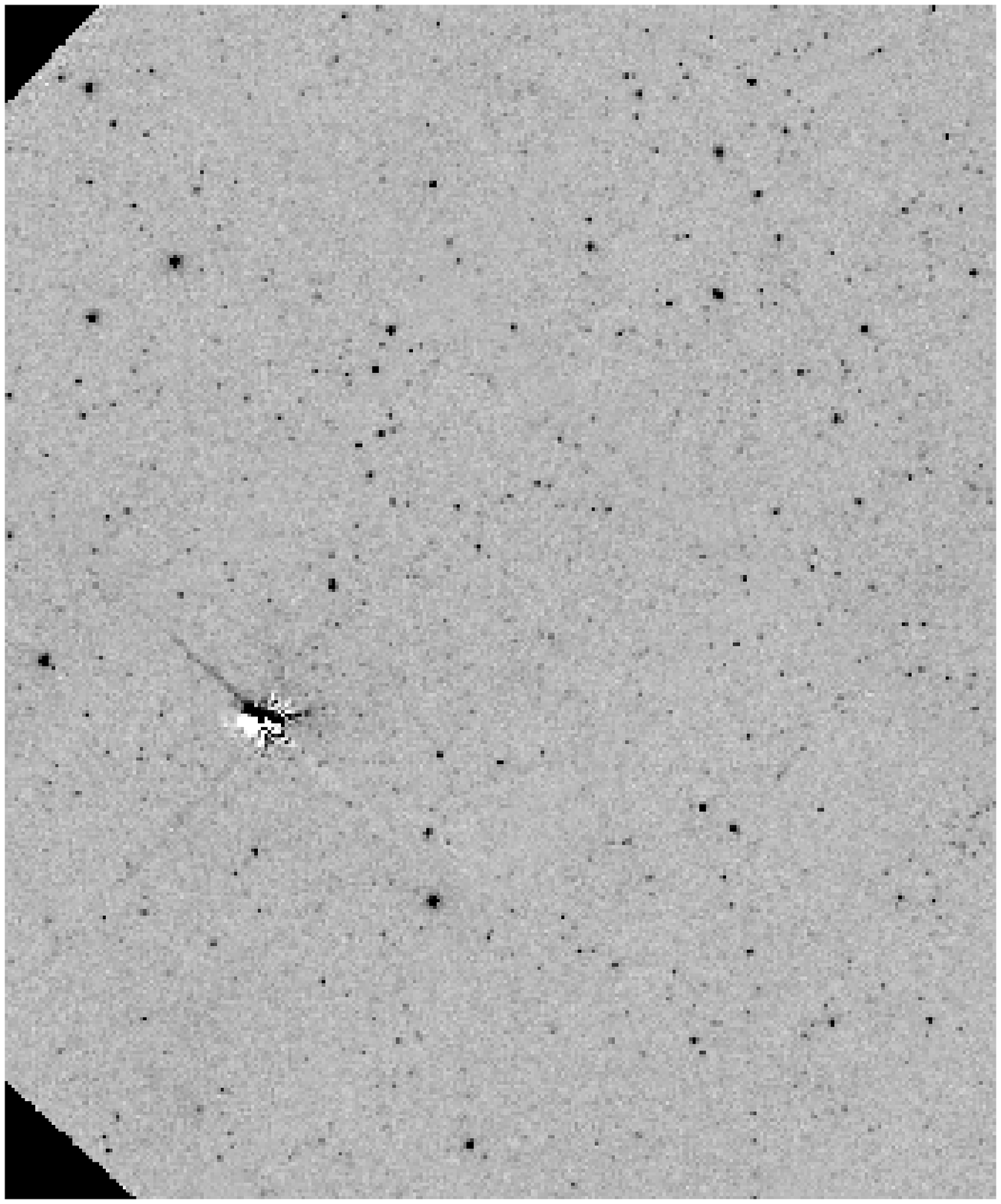}
        \end{minipage}%
    }
    \caption[]{{\bf Partial IRAC 4.5$\mu$m mosaic.} We show a portion of the IRAC 2 mosaic 14.5$^{\prime}$~$\times$~11$^{\prime}$. The images are centered at (13$^{h}$36$^{m}$12$^{s}$,41$^{o}$00$^{\prime}$38$^{\prime\prime}$) showing the star which creates the worst droop. Before droop correction is shown on the left. On the right, is the same mosaic after the droop and background matching corrections. North is up and East is to the left.} \label{droop}
\end{figure*}

There are now two more effects from the bright stars and cosmic rays that we can
remove. The first is to correct for the droop. This a small variation
in the zero-level background in frames where stars
appear. Figure~\ref{drooppl} shows the profile of the negative offset in the flux of the rows along one
frame with a star. Adjacent to the star, the background level drops
(or droops) before ramping up to the level of the mean. We use wavelength filtering to find the location and level of the sharp discontinuity caused by the offending bright source. The resulting filtered profile 
is subtracted from the image. Additional noise caused by subtracting this imperfectly smooth profile is less than the RMS level of the BCD. Finally, the
overall background level for the BCDs in which the bright stars are located are 
artificially higher than the other dithers. Therefore, we replace this high background. We calculate the image median, subtract it and add in the average median value of the frames within 7.5$^{\prime}$ and for which the star does not appear. Figure~\ref{droop}
shows that the correction removes the effect.

The estimates for the contribution from zodiacal light \citep{kel98} and the ISM \citep{sch98} are found in the BCD image header. Both are subtracted from each BCD.

\subsubsection{Superflat} 

In IRAC 5.8 and 8.0$\mu$m there is a gradient within each individual BCD for
which the background level varies smoothly across the frame. To
remove this we construct a superflat field. We divide each BCD by its median value, and then construct the superflat from the median value of all the individual flattened images. The original BCDs are then divided by the smoothed superflat. 

\subsubsection{Stray light}

Several frames in IRAC 3.6$\mu$m of Abell~1763 were subject to stray
light causing false sources. In the couple of cases where the effect
was strong enough to raise the level of the entire frame, we
chose not to include the observation. However, usually the stray light
could be localized and masked. Frames where there was a small (up to
30\%) increase in the background light were renormalized to the same
level as the surrounding frames.

\subsubsection{Final mosaics}

\begin{figure*}
 \subfigure{
 \begin{minipage}[c]{0.5\textwidth}
        \centering
        \includegraphics[width=3.1in,angle=0]{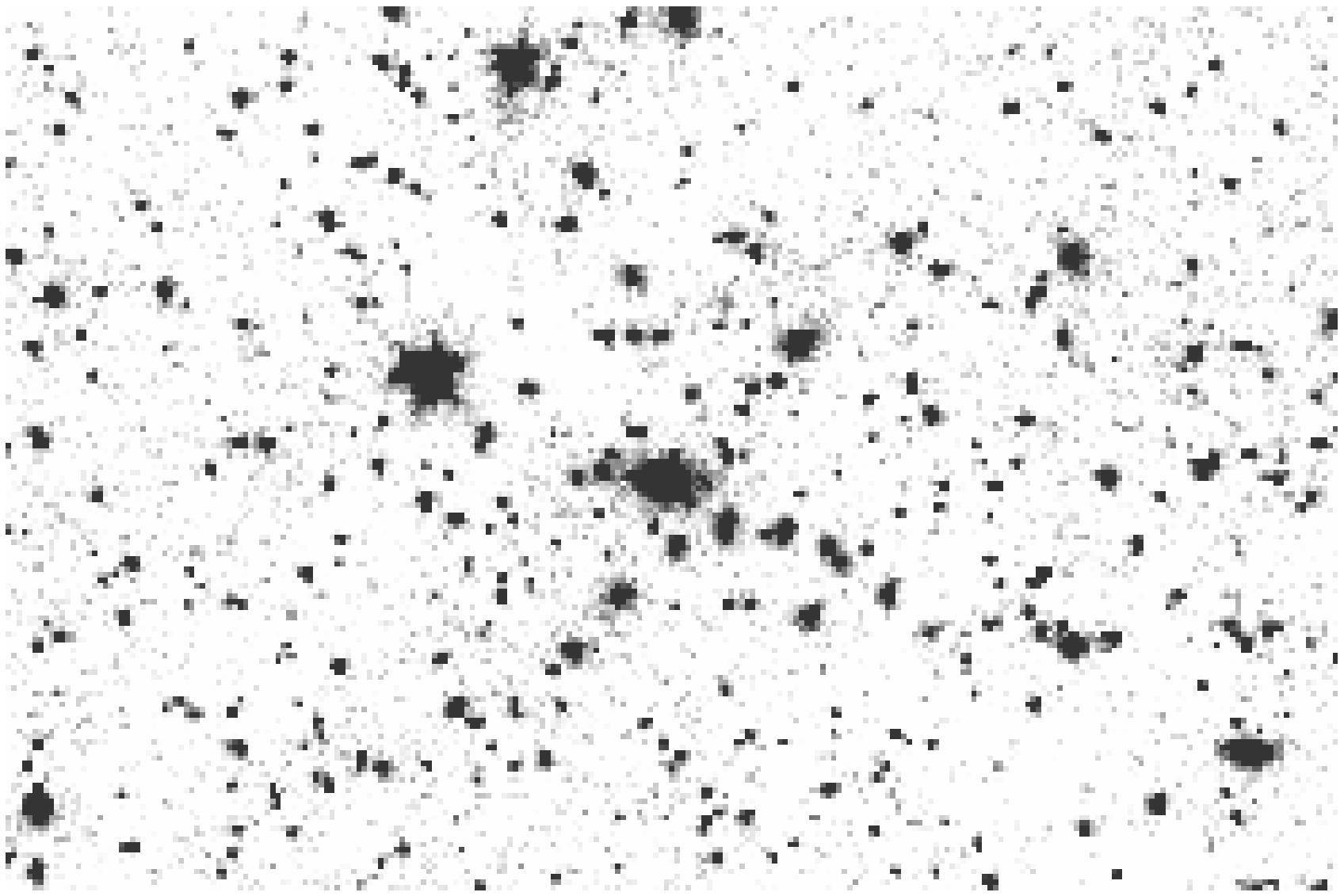}
        \end{minipage}%
    }
    \subfigure{
     \begin{minipage}[c]{0.5\textwidth}
        \centering
        \includegraphics[width=3.1in,angle=0]{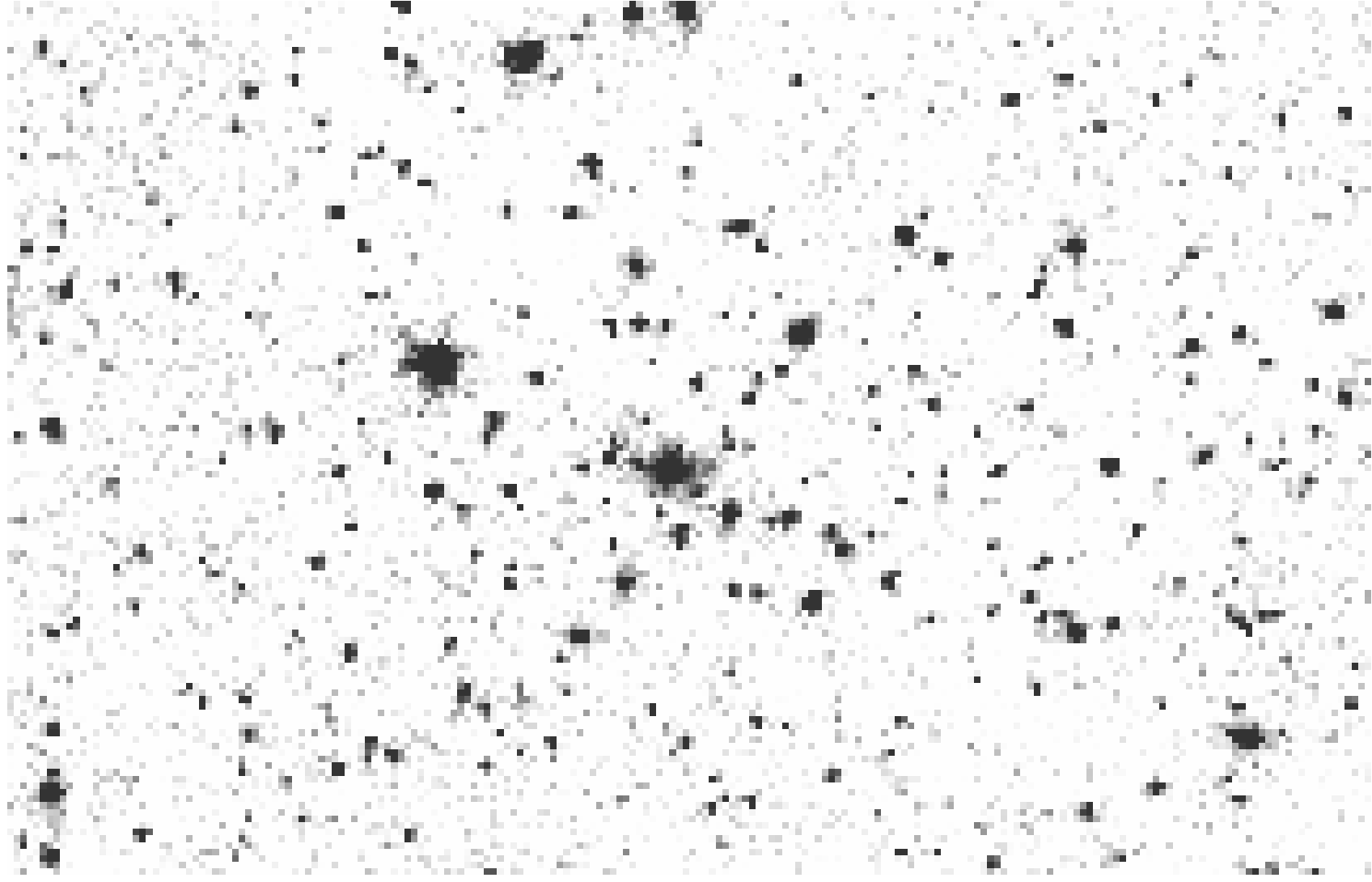}
         \end{minipage}%
    }
 \subfigure{
 \begin{minipage}[c]{0.5\textwidth}
        \centering
        \includegraphics[width=3.1in,angle=0]{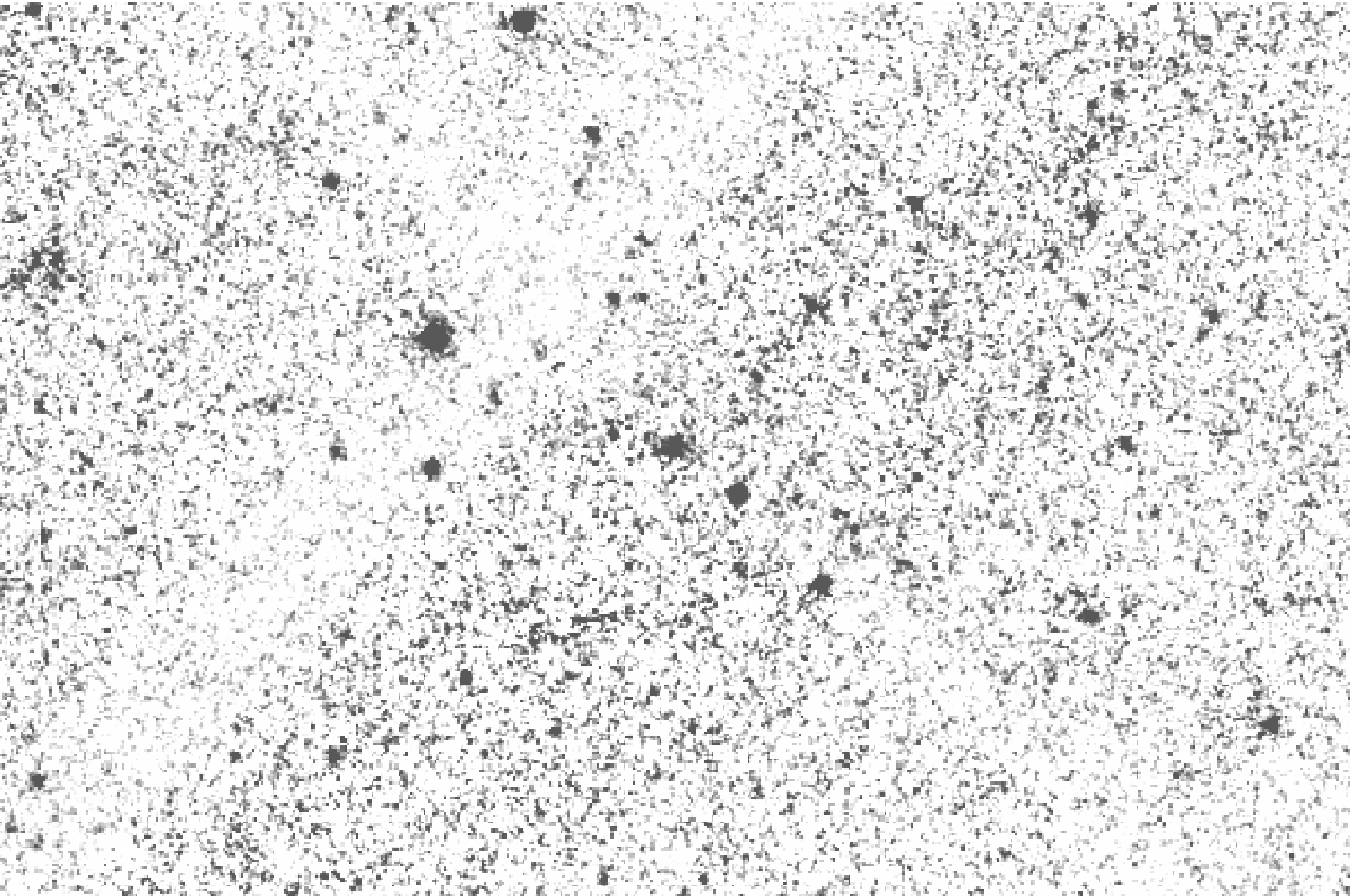}
        \end{minipage}%
    }
 \subfigure{
 \begin{minipage}[c]{0.45\textwidth}
        \centering
        \includegraphics[width=3.1in,angle=0]{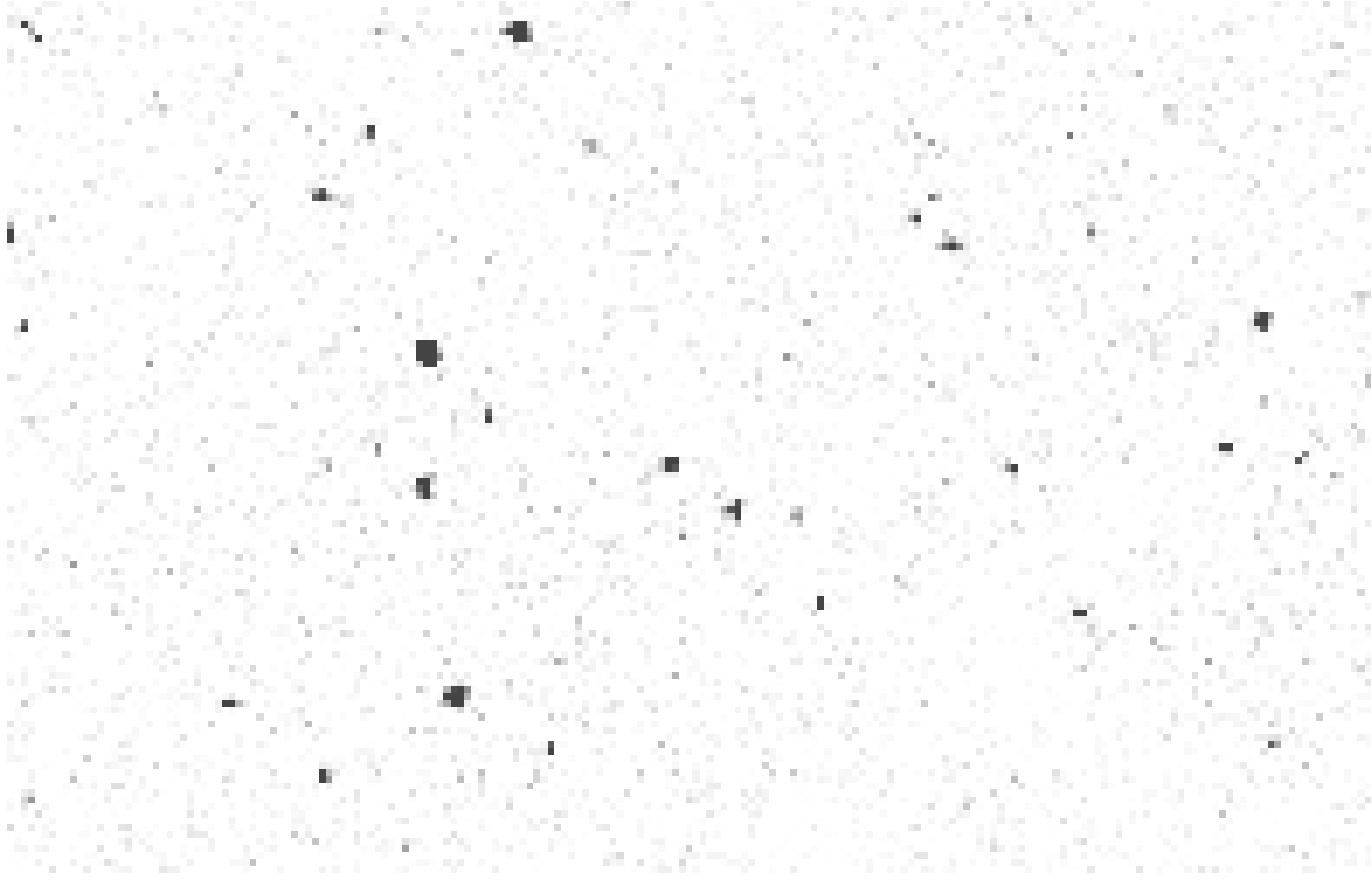}
        \end{minipage}%
    }

    \caption[]{{\bf IRAC images of Abell 1763.}
    The final reduced IRAC images, on the top from left to right: 3.6 and 4.5$\,$$\mu$m, on the bottom from left to right: 5.8 and 8.0$\,$$\mu$m. We show a field of 9.5$^{\prime}$ by 6$^{\prime}$ centered at (13$^{h}$35$^{m}$19$^{s}$,41$^{o}$00$^{\prime}$00$^{\prime\prime}$), near the cluster center. Channel 3, 5.8$\,$$\mu$m, is known to have a lower SNR than the other IRAC channels, and this is evident in our images as large-scale residuals which correspond to the coverage map still appear. North is up and East is to the left.}
    \label{A1763iims}
\end{figure*}

After this rigorous correction for the image defects and the effects
of bright stars, the BCDs are mosaiced together using the script {\it mosaic.pl} in MOPEX (MOsaicer and Point source EXtractor, \citet{mako05,mak05}). The final mosaiced IRAC images have dimensions of 39.2$^{\prime}$$\times$39.2$^{\prime}$ (8.5$\,$Mpc$\times$8.5$\,$Mpc) and are cosmic ray corrected. These final images are shown in Figure~\ref{A1763iims}. The astrometry has been registered to 2MASS. They have been calibrated from DN~s$^{-1}$ to MJy~sr$^{-1}$ with the FLUXCONV  values of 0.1088, 0.1388, 0.5952, and 0.2021 \citep{rea05} for IRAC~3.6, 4.5, 5.8, and 8.0$\,$$\mu$m, respectively.


\subsubsection{Source extraction}

We use SExtractor \citep{ber96} for extracting the IRAC sources. We make no attempt to deblend sources. However, we flag sources with close pairs (within a project separation $<$7$^{\prime\prime}$ from another source at the same wavelength). This will miss any sources that are highly blended, but given the high resolution of the IRAC images, we do not consider this a problem for the cluster galaxies, who's typical size is well above the IRAC spatial resolution. For IRAC~3.6$\mu$m, ~4.5$\mu$m, IRAC~5.8$\mu$m, and IRAC~8.0$\mu$m, at this conservatively large radius, 8, 16, 12, and 17\%, respectively, of the sources are close pairs. We also flag sources associated with stars by matching the position to the SDSS catalogs, Data Release 7. If the IRAC source is within 1$^{\prime\prime}$ of a star as determined by the SDSS star/galaxy separator flag, the IRAC source is flagged as a star. For IRAC ~3.6$\mu$m, ~4.5$\mu$m, IRAC~5.8$\mu$m, and IRAC~8.0$\mu$m, 12\%, 9\%, 6\%, and 2\% of the sources, are flagged as stars, respectively.

We calculate three aperture fluxes using aperture diameters of 4,6, and 12$^{\prime\prime}$.  The aperture corrections we use follow those of the SWIRE technical note \citet{sur05} which lists corrections for radii of 1.9, 2.9 and 5.8$^{\prime\prime}$. Following their work, we divide the aperture fluxes in each band by the corrections at each aperture radius. The corrections for each aperture radius are 0.736, 0.87, and 0.96, for IRAC~3.6$\mu$m; 0.716, 0.87, and 0.95, for IRAC~4.5$\mu$m; 0.606, 0.80, and 0.94, for IRAC~5.8$\mu$m; and 0.543, 0.700, and 0.940 for IRAC~8.0$\mu$m. To estimate the total flux, we also quote Petrosian fluxes which are based on circular apertures whose radius is the average of the azimuthal light profile out to the radius where the local surface brightness is a factor of 0.2 times the mean surface brightness within the radius \citep{pet76}. We use two Petrosian radii and a minimum radius of 3$^{\prime\prime}$ and apply the aperture correction for extended sources as explained on the IRAC calibration webpage~\footnote{Available http://ssc.spitzer.caltech.edu/irac/calib/}. We quote the uncertainty of the fluxes as those given by SExtractor, but with important caveats. These uncertainty values are known to be lower estimates as they only represent the error from Poissonian noise \citep{val09}. However, we include maps of the image RMS which we calculated from the image and coverage maps as described in Section~\ref{Images}.

\subsection{MIPS Data Reduction}

MIPS is composed of three detectors and we reduced the three sets of data using a different procedure for each set. 

\subsubsection{24$\mu$m}

For the 24$\mu$m data we downloaded the BCDs and calibration frames from the {\em Spitzer}
archive using the {\em Spitzer} pipeline version S16.1.0 for which the factor to convert the flux from instrumental units into MJy~sr$^{-1}$ is 0.0454. We used the calibration
frames only to undo the {\em Spitzer} pipeline corrections made to the BCDs, such as  
the jailbar and flat corrections.
We improve on these corrections by using the entire set of frames
in each single Astronomical Observation Request (AOR). The pipeline corrections are
good on average but can depend on the single AOR because of the different
technique of observation (scan instead of photometric mode) and on latencies
from bright objects observed before the AOR.

We reduce the data following the technique used in \citet{fad06}
for the reduction of the {\em Spitzer} First Look Survey data with two main differences.
First, if one examines the 70$\mu$m and 160$\mu$m during the entire observation, the variation of the stims suggests that
this transient is caused by a variation in the response of the detector. This in turn suggests that the long-term transient has been corrected as a multiplicative effect and not as an additive one as in \citet{fad06}. A second difference is in the correction of fast variations of the background coupled with different positions of the cryogenic scan mirror.
We prove that this effect disappears almost completely when the distortion
due to the different light path associated with the different positions of the
cryogenic scan mirror is taken into consideration.

\begin{figure}[t]
   \center
  \epsscale{0.95}
     \plotone{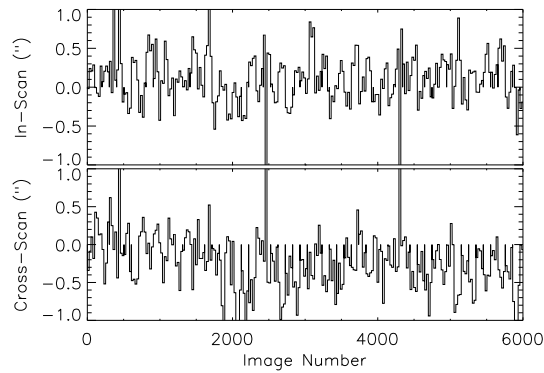}
    \caption{{\bf MIPS position offsets.} The MIPS 24$\mu$m position offsets of each BCD are shown with respect to the SDSS, in arcseconds. 
\label{posoff}}
\end{figure}

We treat the BCDs by first uncorrecting the standard pipeline jailbar and flat field corrections. Subsequently, we apply the following to the BCDs:
(1) our improved jailbar correction computed by masking the bright sources in each frame;
(2) the droop correction apparent in the data rows;
(3) distortion flats, computed from the distortion terms that correct
for the effective pixel area on the sky - this depends on the position
of the cryogenic scan mirror;
(4) superflat correction to take out the real flat from all the BCDs in
each AOR;
(5) flat for each position of the scan mirror to take out the images of the
``dirty spots'' in the cryogenic scan mirror;
(6) removal of zodiacal light;
(7) correction of astrometry offsets by extracting the sources for each
set of 25 consecutive frames and matching to the SDSS catalog. The offsets are similar in both the cross-scan and in-scan directions, we show these in Figure~\ref{posoff}. 

At this point, a first mosaic is obtained and point sources are extracted
with MOPEX. The FWHM of
MIPS 24$\mu$m is 5.9$^{\prime\prime}$, so most of the extragalactic sources
in the field are point sources. In MOPEX we acquire an estimate of the
Point Response Functions (PRF) by taking the mean of $\sim$7 good, isolated, point source
candidates. We then run several of the APEX routines. We first remove the first Airy-ring, which helps constrain the number of false point source detections. We detect point sources on the Airy-ring subtracted image. With this list of 7969 point sources, we then run the flux estimator on the original image. The fluxes are found by fitting the PRF within a radius of 10.3$^{\prime\prime}$. A factor of 1.167 for the aperture correction~\footnote{Available http://ssc.spitzer.caltech.edu/mips/apercorr/} is applied to account for the missing PRF wings.  A color correction of 0.961 is also applied to tie
the photometric system to spectra with $\nu f_{\nu}$=constant (see
MIPS data handbook\footnote{http://ssc.spitzer.caltech.edu/mips/dh},
page 31). 

Some of the point sources may in fact be extended sources and we apply two techniques to extract these, as explained in \citet{fad06}. First, {\em bright\_detect} is run on the original image and the 27 sources with SNR $>$ 20 and ellipticity $>$0.1 are considered as possible extended sources. These are matched to sources in the Optical r$^{\prime}$ image. There are 12 which are clearly extended sources and do not have a close pair. For these, we replace the PRF flux with aperture fluxes calculated out to an extent that includes the entire source. Second, we compare the flux of the central pixel to the flux inside of a 5$^{\prime\prime}$ aperture. Point sources should have a more concentrated flux so that the ratio between the aperture flux to peak flux should be higher in extended sources. There are an additional 16 sources from the point source table for which the ratio is greater than the 3$\sigma$ errorband. However, all of these have very low SNR and so none of the PRF fluxes are replaced.

We run {\it apex\_qa} within MOPEX in order to obtain a residual
image, that is, an image with all the above bright sources (and their first Airy-ring) removed. There exist lower flux extended sources that remain in this residual image and we used SExtractor to measure their source positions. We iterate twice over this process by masking all the sources found in the first iteration, this way, a better background level can be computed. 

To measure the fluxes of these additional 2921 extended sources, we run SExtractor a second time on the residual image (bright sources and their Airy ring's subtracted) by first multiplying by 141.08 to convert the image units of MJy~sr$^{-1}$ to $\mu$Jy (as listed in the MIPS Data Handbook). Petrosian fluxes are calculated using 4 radii and a minimum radius of 5.5$^{\prime\prime}$. We divide by 0.961 to account for the color correction as listed above.

\subsubsection{MIPS 70$\mu$m and 160$\mu$m}

\begin{figure*}
 \subfigure{
     \begin{minipage}[c]{0.45\textwidth}
        \centering
        \includegraphics[width=3in,angle=0]{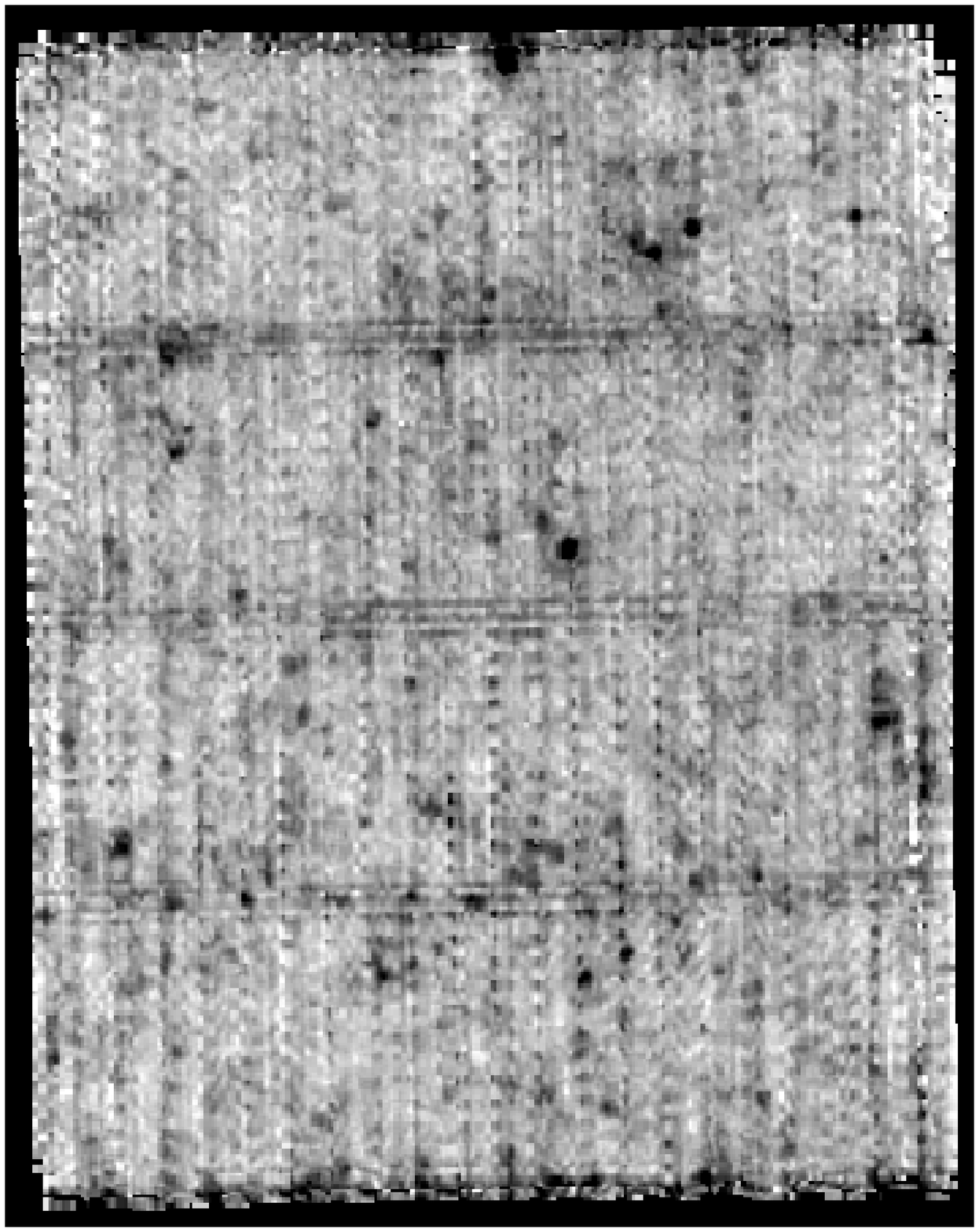}
         \end{minipage}%
    }
 \subfigure{
 \begin{minipage}[c]{0.5\textwidth}
        \centering
        \includegraphics[width=3in,angle=0]{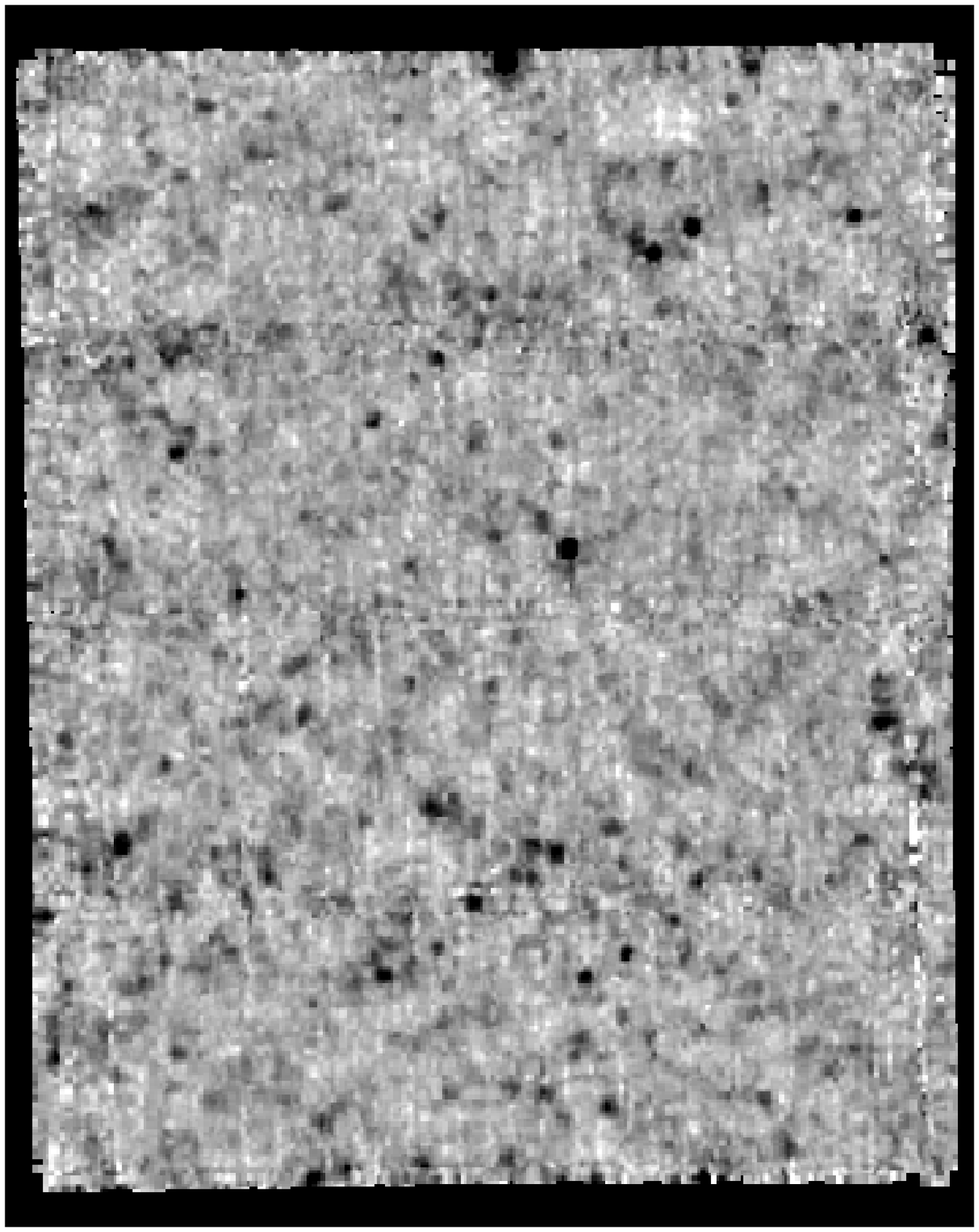}
        \end{minipage}%
    }
    \caption[]{{\bf 160$\mu$m mosaic.} We show the full 160$\mu$m mosaic. The field is 42$^{\prime}$ by 53$^{\prime}$ and centered at (13$^{h}$4$^{m}$46$^{s}$,40$^{o}$52$^{\prime}$04$^{\prime\prime}$). The offline standard pipeline result from GeRT is shown on the left. This figure highlights the much more clean result, on the right, obtained after second pass filtering, cleanup, and ignoring data near the stims. \label{160gert}}
\end{figure*}

\begin{figure*}
 \epsscale{1.5}
    \plotone{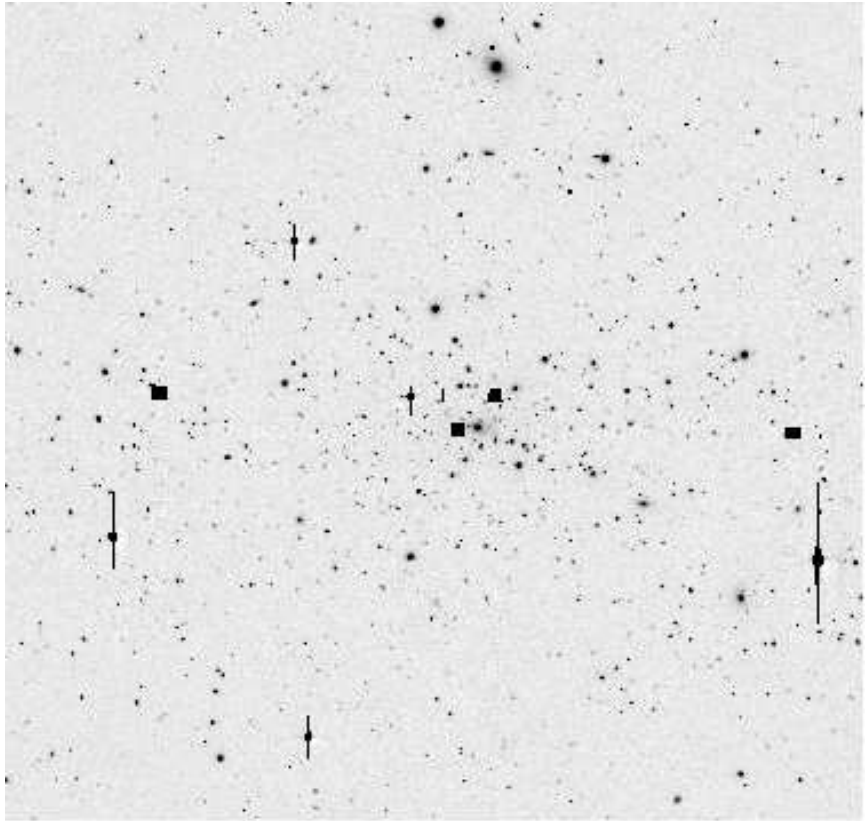}
    \caption{{\bf LFC r$^{\prime}$-band image.} The central $\sim$16$^{\prime}$$\times$16$^{\prime}$ of the optical r$^{\prime}$-band image obtained with the LFC on Palomar. The image is centered at (13$^{h}$35$^{m}$24$^{s}$,41$^{o}$00$^{\prime}$28$^{\prime\prime}$), near the cluster center. The black squares represent areas with no coverage, and the bright stars have been masked to the saturation value of 59100. The associated bad pixel map provided has these pixels flagged with 1. North is up and East is to the left. \label{Opticalr}}
\end{figure*}

In the case of 70$\mu$m and 160$\mu$m observations we downloaded the raw frames (DCEs) from the {\em Spitzer} archive and used the off-line {\em Spitzer} pipeline (GeRT~\footnote{Available at http://ssc.spitzer.caltech.edu/mips/gert/})
to reduce the data.  The conversion factors from instrumental units into MJy~sr$^{-1}$ are 702 and 41.7 for the 70 and 160$\mu$m data, respectively. The pipeline reductions find and calibrate the slope of the data ramp (consecutive data reads). We use second pass filtering to remove streaking in the final image, produced by bright sources. We identify the bright sources in a first pass, and repeat the filtering on the masked image. For the 70$\mu$m data we applied an additional correction to the BCDs to remove the latent stim artifacts by modeling and then removing the effect of the stimflash on the median of the BCDs that follow. In the case of the 160$\mu$m data, we run the {\em cleanup160.pl} script in the GeRT offline package which removes the residual pixel response as a function of time. We also ignore data near the stims when running the mosaicer.

Virtually all of the extragalactic sources at these longer wavelengths are point sources. We mosaic the frames and conduct the source extraction using {\em APEX} with a bottom outlier threshold of 2 and 2.5$\sigma$ and a top threshold of 3 and 2.5$\sigma$ for 70$\mu$m and 160$\mu$m, respectively. We applied no further correction to the astrometry of these longer wavelength sources as the uncertainty of the pointing accuracy of the telescope, $\sim$1$^{\prime\prime}$ is much smaller than the beam widths. For the 70$\mu$m data, we use an aperture radius of 15$^{\prime\prime}$, a color correction of 0.918 and aperture correction of 2.244. For 160$\mu$m we use an aperture radius of 20$^{\prime\prime}$, a color correction of 0.959~\footnote{see p. 31 of the MIPS Data Handbook} and an aperture correction of 3.124. By measuring the noise (RMS) in the mosaiced frames, we arrive at a 5$\sigma$ depth of 6.1$\,$mJy in the 70$\mu$m mosaic, and 85.1$\,$mJy in the 160$\mu$m mosaic. The improvement over using the standard offline GeRT solutions for the 160$\mu$m data is shown in Figure~\ref{160gert}. 

\section{Ancillary ground observations}

In preparation for the {\em Spitzer} and 
spectroscopic follow-up observations, we observed the entire {\em
Spitzer} field with the Large Format Camera (LFC) on the Palomar 200in telescope, obtaining a deep r$^{\prime}$-band image which covers most of the MIPS~24$\mu$m field.
We also obtained further Near-IR imaging which will be used to estimate the
stellar masses of the infrared members of the clusters in a following paper.
Using the Wide-field infrared Camera (WIRC) at the Palomar 200in telescope, we covered the region
including the core of the cluster and part of the filament of galaxies pointing
to the neighbor cluster Abell 1770 (see Figure~\ref{spitzer_cov}). We observed this region in the J, H, and K$_{s}$ bands.

\subsection{Optical r$^{\prime}$-band}

\begin{figure}[h]
 \epsscale{0.9}
    \plotone{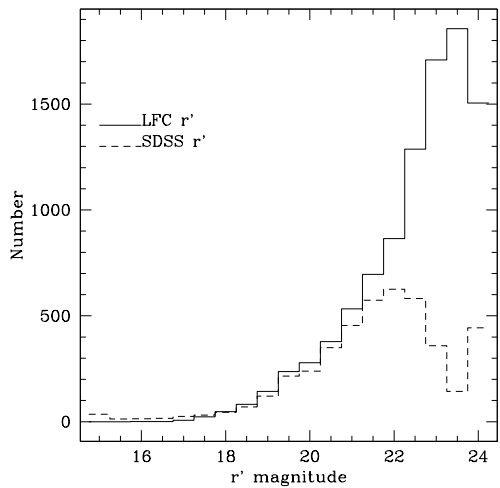}
    \caption{{\bf LFC r$^{\prime}$ catalog depth.}  The histogram shows the number of source counts from the SDSS (dashed line) and our LFC magnitudes (solid line). It is clear that our r$^{\prime}$ magnitudes are deeper than the SDSS by $\sim$2~mag. \label{sdsscomp}}
\end{figure}

At the time of planning the {\em Spitzer} observations, there were no publicly available r$^{\prime}$-band data of this cluster. Although SDSS r$^{\prime}$ data has subsequently been observed and released, we include our LFC images and photometry as it is $\sim$2 mag deeper than the SDSS and because the spatial resolution is higher by a factor of two. To reduce this data, the dark subtraction and flat field corrections were applied in the standard fashion. Bright Tycho stars, as well as regions without coverage, are flagged in the image with the saturation value of 59100. The photometric catalogs were produced with SExtractor using an aperture diameter of 3.5$^{\prime\prime}$, as well as Petrosian magnitudes. The astrometry and photometry are calibrated to the SDSS r$^{\prime}$ positions and Petrosian magnitudes. They are not corrected for Galactic extinction, however this field was chosen in part because of its particularly low extinction level of E(B-V)~=~0.009~mag \citep{sch98}. The image is shown in Figure~\ref{Opticalr}. By calculating aperture magnitudes over areas of the image, free of sources, we derive an average limiting magnitude of 25.5.

\begin{figure*}
 \epsscale{1.2}
    \plotone{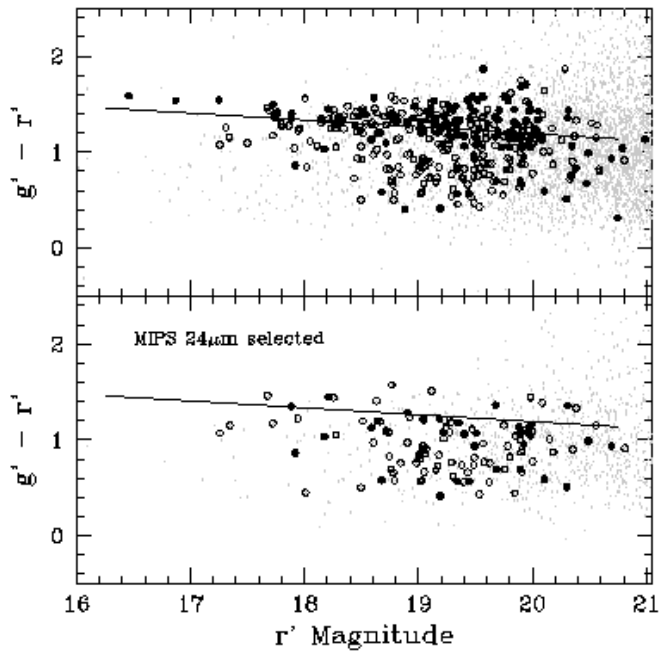}
    \caption{{\bf Optical color magnitude relations}  The optical color magnitude diagram is shown on the top panel and derived using SDSS Petrosian magnitudes. The grey dots are all SDSS galaxies within the field of the supercluster. The open grey circles represent galaxies with confirmed spectroscopic redshifts between z=0.20-0.24. The filled black circles are spectroscopic members within 7.5$^{\prime}$ of the cluster center (taken to be the location of the BCG). The bottom panel shows the same, but only for the MIPS 24$\mu$m selected sources where the MIPS fluxes have S/N$>$5. A linear fit through the cluster galaxies has been calculated on the top panel and applied to the bottom panel showing most MIPS galaxies in the blue portion of the color magnitude relation. \label{optcolors}}
\end{figure*}

We use the SDSS not only to calibrate the LFC positions and photometry, but also to supplement our catalog with photometric points in u$^{\prime}$,g$^{\prime}$,i$^{\prime}$,z$^{\prime}$, and r$^{\prime}$ where LFC photometry does not exist. The SDSS magnitudes and errors we quote are the Petrosian magnitudes taken directly from Data Release~7, available on the SDSS website~\footnote{http://www.sdss.org/dr7/}. For the purposes of comparing to the LFC magnitudes, we match the SDSS magnitudes to the LFC source with the smallest separation in radius, neglecting close pairs (within a distance of 2$^{\prime\prime}$).

We present Figure~\ref{sdsscomp} which shows a histogram of the SDSS r$^{\prime}$ magnitudes along with a histogram of our LFC r$^{\prime}$ magnitudes to show the depth of our magnitudes compared to the SDSS. Both histograms are generated using sources within an area of 25 square arcminutes centered on the brightest cluster galaxy. 

Figure~\ref{optcolors} shows a color magnitude diagram using the SDSS Petrosian magnitudes. Overlaying galaxies with known spectroscopic redshifts, the red sequence is prominent. We also mark the MIPS 24$\mu$m emitting galaxies. Notice how the infrared-selection preferentially finds the star-forming galaxies of the blue cloud, many of which are further than 7.5$^{\prime}$ away from the cluster BCG.

\subsection{Near-IR}

We used the WIRC reduction software of Tom Jarrett~\footnote{http://spider.ipac.caltech.edu/staff/jarrett/wirc/jarrett.html} to reduce the data of all three bands. First, a median dark frame is subtracted for all frames followed by an application of flux-non linearity terms. The Near-IR sky varies rapidly, therefore the data are processed in sets of no more than 12 frames. A median sky is calculated for each set, and subtracted from each frame. Each frame is then flat fielded. 

\begin{figure}
 \epsscale{1.1}
    \plotone{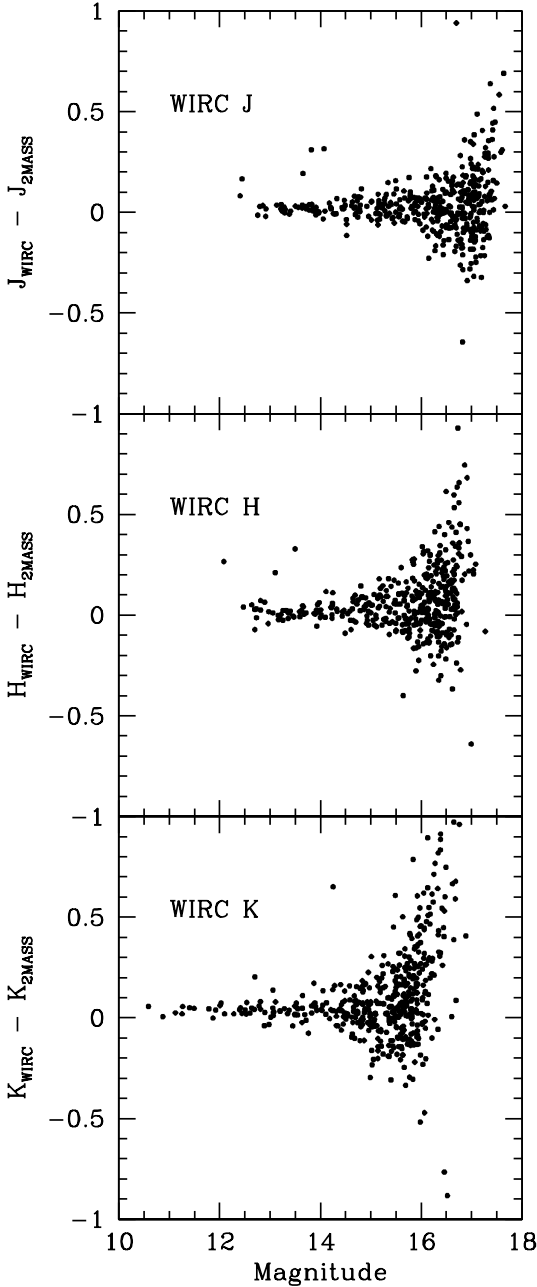}
    \caption{{\bf WIRC calibration.} A comparison of the WIRC aperture magnitudes to the best 2MASS aperture magnitudes is executed in order to produce refined calibration. The points show the difference in the calibrated WIRC magnitude with the 2MASS magnitude as a function of the WIRC magnitude. Each panel shows $\sim$450 2MASS sources for which there exists a WIRC counterpart free of a close pair.  \label{Wirccal}}
\end{figure}

\nocite{skr06}
Each time that WIRC is installed on Palomar, there is a small rotational offset of about 1.5$^{o}$. To get the best astrometry, we can measure the exact offset using a list of known point sources and comparing their known positions from the Two Micron All Sky
Survey (Skrutskie et al. 2006; hereafter, 2MASS) which we obtain from the {\em NASA/IPAC Infrared Science Archive (IRSA)/Gator} webtool. We apply the rotational offset to the data. Using this list of point sources with known positions, we then
examine every image by eye to ensure good astrometry corrections,
which are tweaked by hand if necessary. 

\begin{figure*}
 \subfigure{
 \begin{minipage}[l]{0.9\textwidth}
        \centering
        \label{J}
        \includegraphics[width=4in,angle=0]{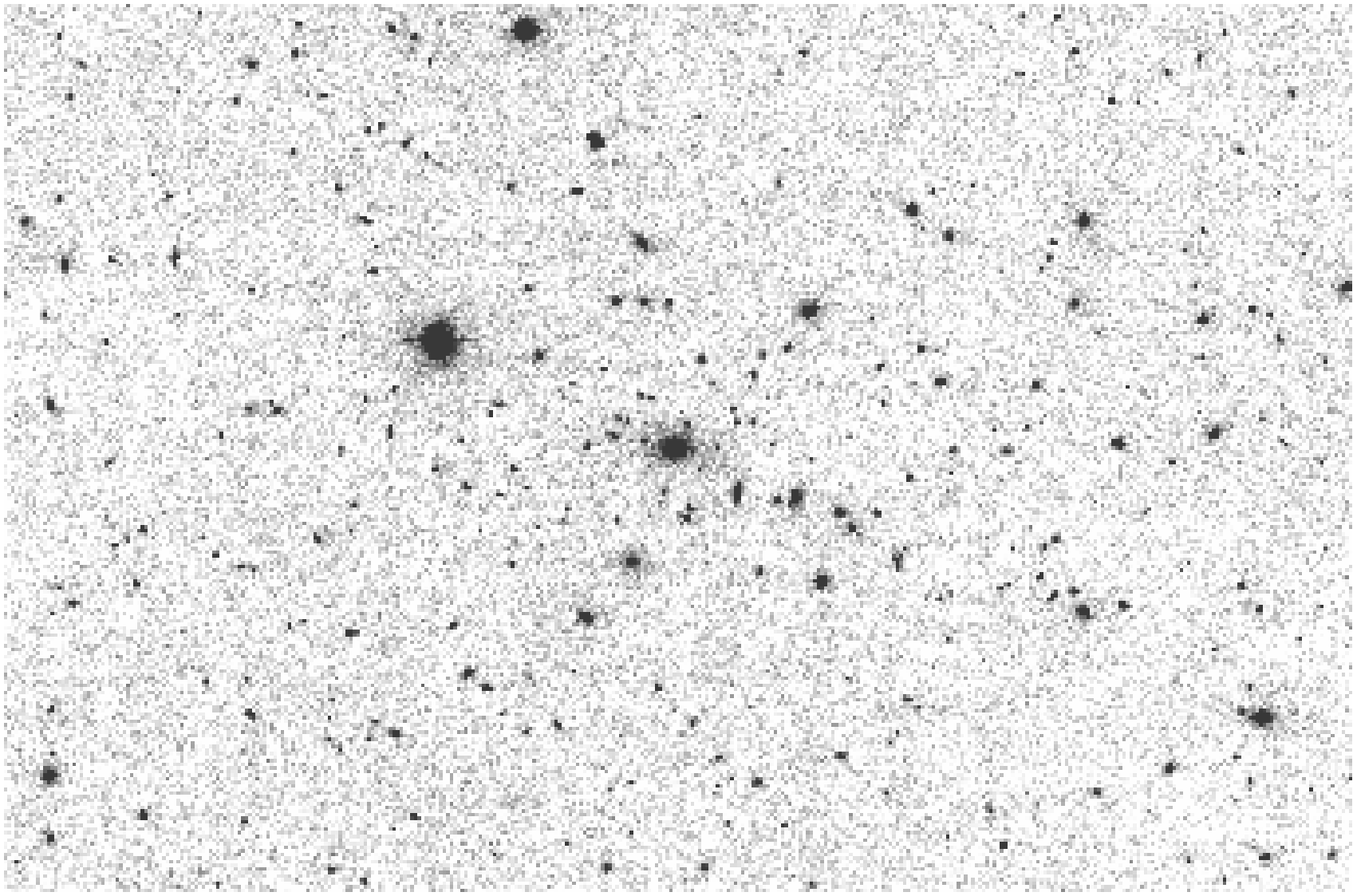}
        \end{minipage}%
    }
    \subfigure{
     \begin{minipage}[l]{0.9\textwidth}
        \centering
        \label{H}
        \includegraphics[width=4in,angle=0]{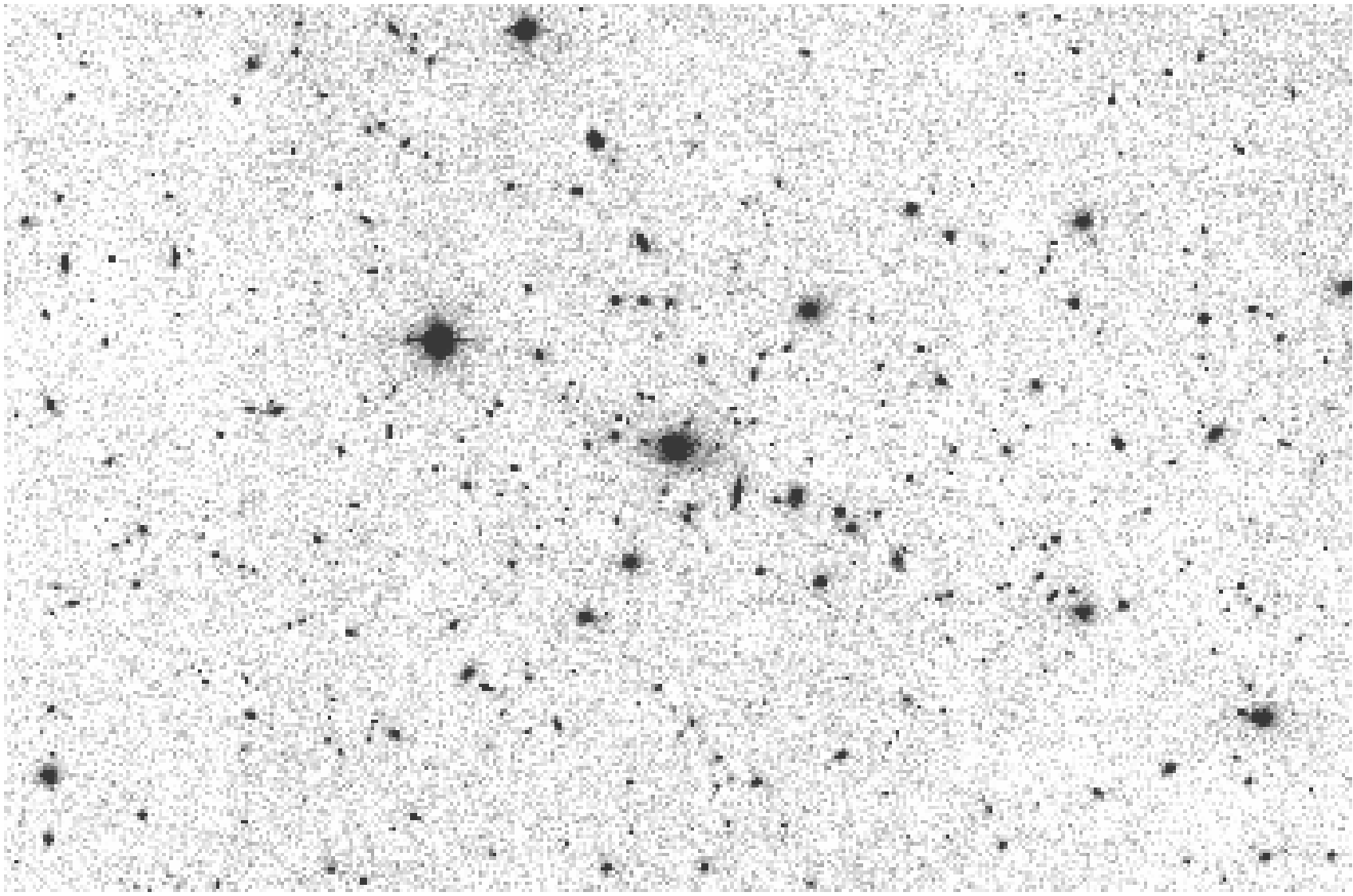}
         \end{minipage}%
    }
 \subfigure{
 \begin{minipage}[l]{0.9\textwidth}
        \centering
        \label{K}
        \includegraphics[width=4in,angle=0]{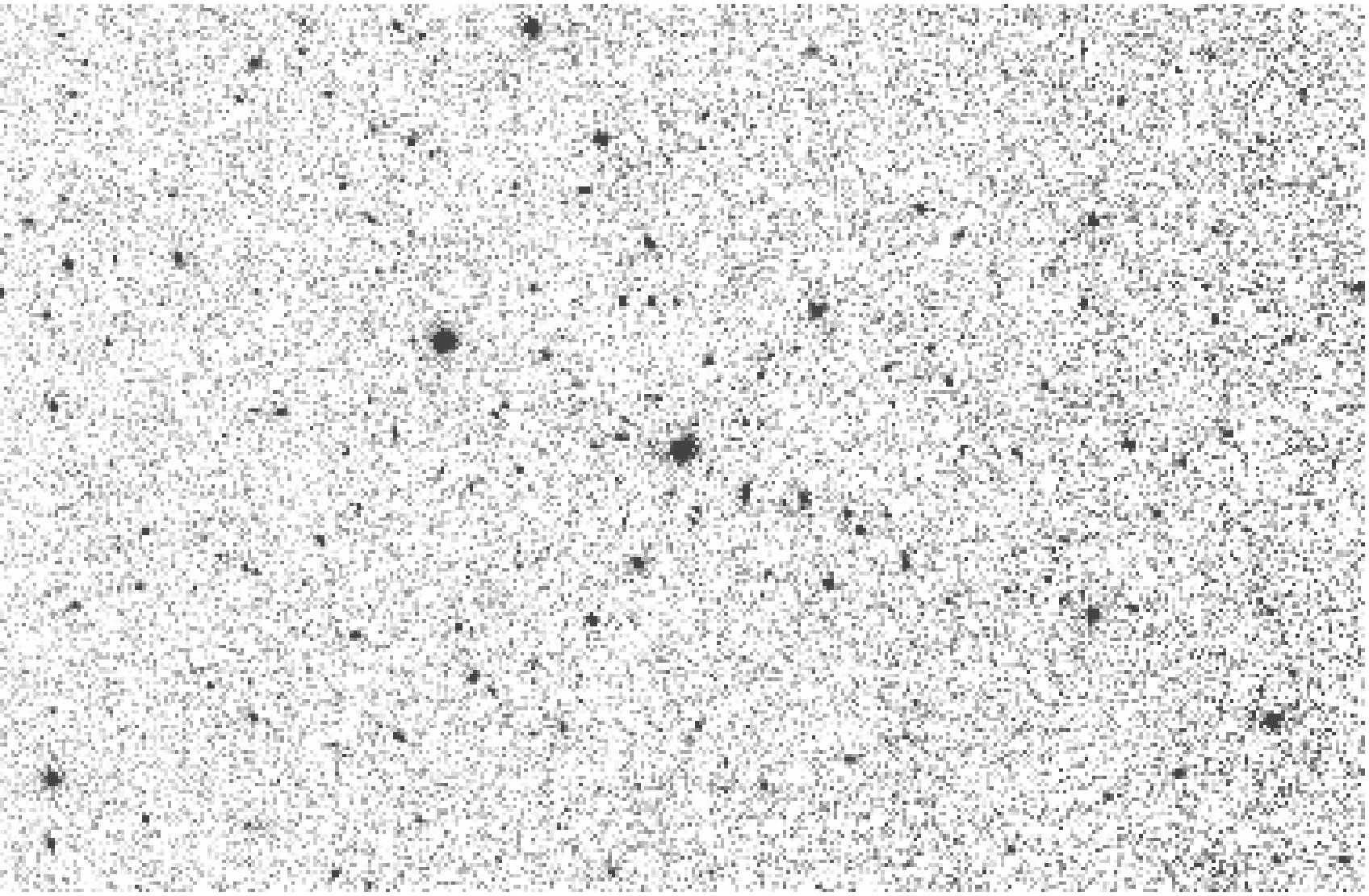}
        \end{minipage}%
    }
    \caption[]{{\bf WIRC images of Abell 1763.} We show the central 9.5$^{\prime}$ by 6$^{\prime}$ of the final mosaics of WIRC J, H and K$_{s}$ band imaging displayed respectively from top to bottom. The frames are centered on the brightest cluster galaxy at (13$^{h}$35$^{m}$20$^{s}$,41$^{o}$00$^{\prime}$03$^{\prime\prime}$). North is up and East is to the left, all of the images are shown in at a similar stretch.}\label{jhk}
\end{figure*}

The background level of each frame is calculated and removed. A dark frame is subtracted from each image and the flux bias correction is run, which applies the non-linearity coefficients. The individual dithers are calibrated using stars from 2MASS with magnitudes from 11-14 in 7$^{\prime\prime}$ radius apertures. Frames where the photometric uncertainty, or where the background is high with respect to 2MASS are discarded. We remove $\sim$20 frames from $\sim$200 taken in each of J, H, and K. We use {\it SWarp} \citep{ber07} to mosaic the frames together and then do a final flux calibration on the mosaic relative to the same 2MASS catalog. This time the best 2MASS point source total magnitudes are compared to the Petrosian magnitudes obtained from SExtractor, which are a good estimate of the total magnitudes. This refines the calibration by +0.07, -0.05, and +0.09 magnitudes in J, H, and K$_{s}$, respectively, as shown in Figure~\ref{Wirccal}. The red tail at faint fluxes is created by 2MASS detections whose fluxes are boosted by a positive noise bias. The photometry of the final mosaiced images is excellent and we calculate an uncertainty of 0.047, 0.041, and 0.042 magnitudes in J, H, and K$_{s}$, respectively. Figure~\ref{Wirccal} shows that at the faintest magnitude level, the dispersion of the difference is $\sim$4\%. Assuming a similar error for the WIRC and 2MASS photometry, then the WIRC error is $\sim$4\%/$\sqrt{2}$$\sim$3\%. In fact most of the scatter at the faint level likely occurs because of the proximity to the detection limit in the 2MASS data. For K$<$15, the WIRC error is $\sim$2\%/$\sqrt{2}$$\sim$1.4\%.

The final J, H and K$_{s}$ mosaics are shown in Figure~\ref{jhk}. The photometry was performed over the final mosaic.

\begin{figure}
 \epsscale{1.1}
    \plotone{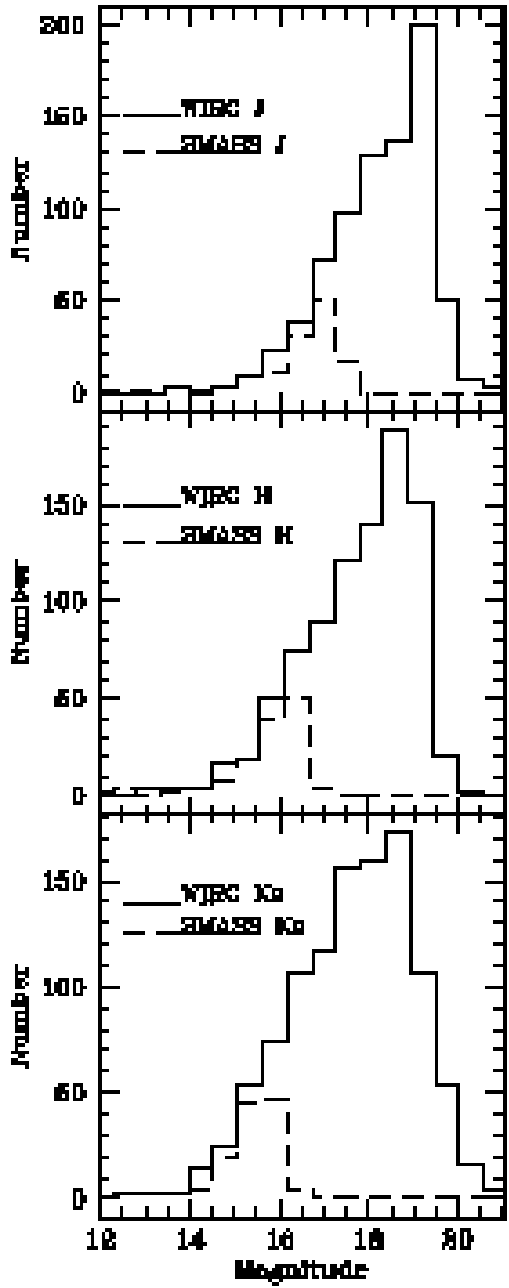}
    \caption{{\bf WIRC catalog depth.} The histograms show the number of source counts from the 2MASS (dashed line) and our WIRC magnitudes (solid line), for J, H and K$_{s}$ from top to bottom. The histograms are made using only the central 10$^{\prime}$~$\times$~10$^{\prime}$ of the mosaic and include only WIRC magnitudes with a SNR $>$ 10 for J and K$_{s}$, and SNR $>$ 15 for H. It is clear that our WIRC magnitudes are deeper than the 2MASS by $\sim$3~mag. \label{2masscompj}}
\end{figure}

SExtractor is used to find aperture magnitudes within a 3.5$^{\prime\prime}$ diameter and Petrosian magnitudes for all sources. To ensure we extract the small as well as very extended objects that are apparent in the WIRC images, we use a {\em det\_minarea} of 5 and {\em det\_minthresh} of 2.0. To ensure we find the smaller objects that often surround larger objects (ex. near the BCG) we use a {\em deblend\_mincont} of 0.00005.

Our WIRC catalogs are not the deepest observations of cluster galaxies (\citet{wuy08}, for example, go down to K$_{s}$$\sim$24.3). However, for our purpose which is to study the galaxies down to $\sim$M$^{*}$-2 in the Abell~1763 cluster and filament, they suffice. The K$_{s}$* of cluster galaxies at z=0.2 is 15.6 \citep{dep03}, and typical colors of late type galaxies are J-K$_{s}$ = 1.7 and H-K$_{s}$ = 0.8 \citep{fuk95}, therefore our Near-IR catalogs need to reach J$\sim$19.5, H$\sim$19, and K$_{s}$$\sim$18. Figure~\ref{2masscompj} shows that we indeed accomplish this. Comparing with the depth of the 2MASS catalogs, we note that our WIRC magnitudes are deeper than 2MASS by $\sim$3~mag.

\begin{figure*}
 \epsscale{1.2}
    \plotone{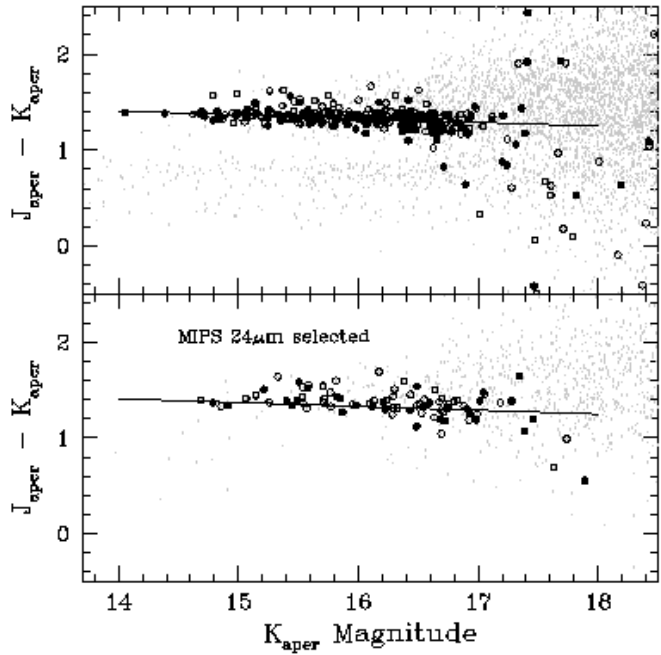}
    \caption{{\bf Near-IR color magnitude relations}  The Near-IR color magnitude diagram is shown on the top panel, derived using aperture magnitudes. The grey dots are all WIRC galaxies within the field of the supercluster. The open grey circles represent galaxies with confirmed spectroscopic redshifts between z=0.20-0.24. The filled black circles are spectroscopic members within 7.5$^{\prime}$ of the cluster center (taken to be the location of the BCG). The bottom panel shows the same, but only for the MIPS 24$\mu$m selected sources, where the MIPS fluxes have S/N$>$5. A linear fit through the cluster galaxies has been calculated on the top panel and applied to the bottom panel. \label{nircolors}}
\end{figure*}

As the catalog we present includes aperture magnitudes of the same diameter (3.5$^{\prime\prime}$) in r$^{\prime}$, J, H, and K$_{s}$, and as the resolution in these mosaics is all of order 1.5$^{\prime\prime}$, we can examine the colors of the cluster galaxies (and the MIPS-selected galaxies). Figure~\ref{nircolors} shows that the Near-IR color magnitude relation reveals a tighter red sequence for the cluster than the optical color magnitude diagrams presented in the previous section. The Near-IR colors are driven by old stars, so there is less scatter due to younger stellar populations. More work on the colors of the Abell~1763 cluster and filament galaxies will be presented in subsequent papers.

\section{Catalogs}

We compile two cluster catalogs which we also release to the community online via {\em IRSA}. The first gives 24$\mu$m source fluxes and their associated magnitudes in the five optical SDSS bands, in the deeper LFC r$^{\prime}$, in the three WIRC bands, and their fluxes in the four IRAC bands, MIPS~70$\mu$m, and MIPS~160$\mu$m. The second catalog presents 70$\mu$m source fluxes in the regions where there is no 24$\mu$m coverage. We also include the magnitudes from the five optical SDSS bands.

\subsection{24$\mu$m catalog}

There are two flags based on the 24$\mu$m sources which are also included. The first flags close pairs of  24$\mu$m sources. We make no effort to deblend the 24$\mu$m sources beyond what has been applied by APEX routines, however we do provide a flag for sources that have at least one other source within 6.1$^{\prime\prime}$. Only 5\% of the sources have close pairs. The MIPS 24$\mu$m FHWM is quite large and it is reasonable to assume that closer pairs may exist that we can not resolve as they would be highly blended. If we assume that the galaxy source counts in the IRAC bands are related to those in MIPS 24$\mu$m we can deduce that blending is not of great concern as the number of close pairs, over the same 6.1$^{\prime\prime}$ radius, in the much higher resolution IRAC 3.6$\mu$m image is similar at 4\%. The second flag we list marks the MIPS 24$\mu$m sources associated with stars as defined by the SDSS star/galaxy separator. The catalog also lists the flux of the associated sources from the other wavebands we studied. 

\begin{figure}
 \epsscale{0.9}
    \plotone{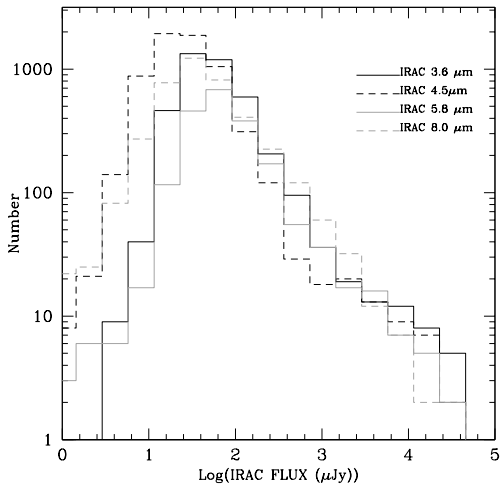}
    \caption{{\bf IRAC fluxes.}  The histogram shows the number of IRAC sources at each flux level. Close pairs and bright SDSS stars have been removed from the plot. \label{iracmag}}
\end{figure}

For all wavelength bands short  of 70$\mu$m, the sources have been associated as those with the highest reliability. We calculate the likelihood of the candidate counterpart to be a real association rather than a background source, as described in \citet{sut92} and \citet{cil03}. The reliability is the likelihood ratio normalized over all possible candidates and accounts for the separation as well as the relative magnitude distribution of the two data sets. We consider only associations within a 5$^{\prime\prime}$ radius of the MIPS 24$\mu$m source, and include only associations which have a likelihood ratio greater than 0.2 (see Fadda et al. (2006) for a more detailed description). For the longer wavelengths, there are fewer sources and the resolution is lower than that of the 24$\mu$m catalog, thus we simply choose the candidate with the shortest projected distance. If a 24$\mu$m source does not have a flux associated to a particular band, it is flagged with -999. 

\begin{figure}
 \epsscale{0.9}
    \plotone{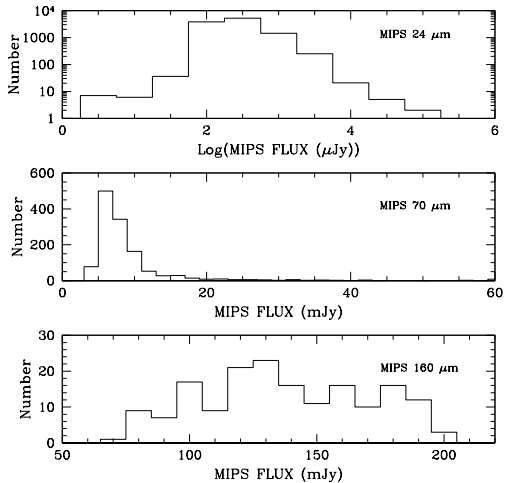}
    \caption{{\bf MIPS fluxes.}  The histograms show the number of MIPS sources at each flux level. MIPS 24$\mu$m is shown at the top, MIPS 70$\mu$m at the center, and MIPS 160$\mu$m on the bottom. \label{mipsmag}}
\end{figure}

Figures~\ref{iracmag} and \ref{mipsmag} show the histogram of source fluxes in the IRAC and MIPS mosaics.

Table~\ref{cat24} describes the machine readable 24$\mu$m source catalog. The first column lists the columns of the catalog that correspond to each entry. The second column lists the format of those columns, whether it is an integer or float, and the number of characters. The third column gives a short description of each entry. We now describe column~3 of this table in a more detail. The first row simply gives a number to the catalog row. The second two rows give the J2000 RA and DEC position in decimal degrees. Subsequently we list the flux and error of the MIPS 24$\mu$m source in $\mu$Jy. The SDSS u$^{\prime}$, g$^{\prime}$, r$^{\prime}$,i$^{\prime}$,z$^{\prime}$ Petrosian magnitudes and their errors are listed, followed by 2MASS J, H, and K total magnitudes and errors. Our own deeper r$^{\prime}$  and WIRC J, H, and K$_{s}$ aperture and Petrosian magnitudes follow with the errors quoted as those from SExtractor. The SDSS magnitudes are given in their native AB magnitude system, as is the LFC r$^{\prime}$ magnitude. The J, H and K$_{s}$ magnitudes are in the Vega magnitude system as that is the native magnitude system of 2MASS, to which they have been calibrated. We list the fluxes in $\mu$Jy for the longer wavelengths: first, the three aperture and Petrosian fluxes in the four IRAC bands and their errors, then the MIPS 70$\mu$m PSF flux and error, and finally the 160$\mu$m PSF flux and error. Finally we list several flags. The first two are given a value of 1 for true. The first flag states whether the MIPS 24$\mu$m source is in a close pair, the second whether it is a star by the SDSS definition. The next eight values list the stellarity index of, respectively, the IRAC 3.6 - 8.0$\mu$m sources, followed by the WIRC J, H, K$_{s}$, and LFC r$^{\prime}$ sources (all calculated in SExtractor). The final 14 flags are the close pair flags for the IRAC and WIRC bands followed by the star/galaxy flags for the IRAC and WIRC bands.

\begin{deluxetable}{lcl}
\tabletypesize{\scriptsize}
\tablewidth{0pt}
\tablecaption{MIPS 24 $\mu$m Source Catalog Columns \label{cat24} }
\tablehead{\colhead{Column} & \colhead{Format} & \colhead{Description}}
\startdata
1-6 & i6 & Catalog Number\\
8-22 &f14.6 &RA (J2000)\\
24-38 &f14.6 &DEC (J2000)\\
40-49 &f10.3 &MIPS24 ($\mu$Jy)\\
51-60 &f10.3 &MIPS24e ($\mu$Jy)\\
62-71 &f10.3 &u$^{\prime}$ (mag)\\
73-82 &f10.3 &u$^{\prime}$e (mag)\\
84-93 &f10.3 &g$^{\prime}$ (mag)\\
95-104 &f10.3 &g$^{\prime}$e (mag)\\
106-115 &f10.3 &r$^{\prime}$ (mag)\\
117-126 &f10.3 &r$^{\prime}$e (mag)\\
128-137& f10.3 &i$^{\prime}$ (mag)\\
139-148& f10.3 &i$^{\prime}$e (mag)\\
150-159 &f10.3 &z$^{\prime}$ (mag)\\
161-170 &f10.3 &z$^{\prime}$e (mag)\\
172-205 & 3xf10.3 & 2MASS J,H,K (mag) \\
207-146 & 4xf10.3 &LFC r$^{\prime}$,r$^{\prime}$e ap 3.5$^{\prime\prime}$,Petrosian (mag)\\
148-287 & 4xf10.3 &J,Je ap 3.5$^{\prime\prime}$,Petrosian (mag)\\
289-328 & 4xf10.3 &H,He ap 3.5$^{\prime\prime}$,Petrosian (mag)\\
330-369 & 4xf10.3 &K$_{s}$,K$_{s}$e ap,Petrosian (mag)\\
371-450 & 8xf10.2 &IRAC1,IRAC1e ap 4$^{\prime\prime}$,ap 6$^{\prime\prime}$,ap 12$^{\prime\prime}$,Petrosian ($\mu$Jy)\\
452-531 & 8xf10.2 &IRAC2,IRAC2e ap 4$^{\prime\prime}$,ap 6$^{\prime\prime}$,ap 12$^{\prime\prime}$,Petrosian ($\mu$Jy)\\
533-612 & 8xf10.2 &IRAC3,IRAC3e ap 4$^{\prime\prime}$,ap 6$^{\prime\prime}$,ap 12$^{\prime\prime}$,Petrosian ($\mu$Jy)\\
614-693 & 8xf10.2 &IRAC4,IRAC4e ap 4$^{\prime\prime}$,ap 6$^{\prime\prime}$,ap 12$^{\prime\prime}$,Petrosian ($\mu$Jy)\\
695-704 &f10.2 &MIPS70 ($\mu$Jy)\\
706-715 &f10.2 &MIPS70e ($\mu$Jy)\\
717-726 &f10.2 &MIPS160 ($\mu$Jy)\\
728-737 &f10.2 &MIPS160e ($\mu$Jy)\\
739-740 &i2 & MIPS24 separation flag \\
742-743 &i2 & MIPS24 star/galaxy flag \\
745-800& 8xf7.3 & IRAC1-4,J,H,K,LFC r$^{\prime}$ stellarity index\\
802-857& 8xf7.3 & IRAC1-4,J,H,K close pair flag\\
859-914& 8xf7.3 & IRAC1-4,J,H,K star/galaxy flag\\
\enddata
\end{deluxetable}

\subsection{70$\mu$ catalog}

Approximately thirty percent of the 70$\mu$m image is not covered by the 24$\mu$m image (nor the IRAC, MIPS 160$\mu$m, or WIRC fields of view). We construct a separate catalog for these 733 sources. Table~\ref{cat70} gives the column headings and descriptions for the machine readable catalog. The first column in this table lists the columns of the catalog that correspond to the entry. The second column lists the format of those columns, whether the value is an integer or float, and the number of characters. The third gives a short description. The first column of the machine readable catalog gives the catalog number, the second two give the J2000 RA and DEC in decimal degrees. The third and fourth columns give the MIPS 70$\mu$m flux and error in mJy. The next 10 columns give the SDSS u$^{\prime}$, g$^{\prime}$, r$^{\prime}$, i$^{\prime}$, z$^{\prime}$ magnitudes and their errors in the AB magnitude system. We matched the 70$\mu$m source catalog to the 2MASS catalog, but do not include the NIR magnitudes as only $\sim$7\% of the 70$\mu$m sources have 2MASS associations.

\begin{deluxetable}{lcl}
\tabletypesize{\scriptsize}
\tablewidth{0pt}
\tablecaption{MIPS 70$\mu$m Source Catalog Columns\label{cat70}}
\tablehead{\colhead{Column} & \colhead{Format} & \colhead{Description}}
\startdata
1-6 & i6 & Catalog Number\\
8-22 &f14.6 &RA (J2000)\\
24-38 &f14.6 &DEC (J2000)\\
40-49 &f10.3 &MIPS70 (mJy)\\
51-60 &f10.3 &MIPS70e (mJy)\\
62-71 &f10.3 &u$^{\prime}$ (mag)\\
73-82 &f10.3 &u$^{\prime}$e (mag)\\
84-93 &f10.3 &g$^{\prime}$ (mag)\\
95-104& f10.3 &g$^{\prime}$e (mag)\\
106-115 &f10.3 &r$^{\prime}$ (mag)\\
117-126 &f10.3 &r$^{\prime}$e (mag)\\
128-137 &f10.3 &i$^{\prime}$ (mag)\\
139-148 &f10.3 &i$^{\prime}$e (mag)\\
150-159& f10.3 &z$^{\prime}$ (mag)\\
161-170 &f10.3 &z$^{\prime}$e (mag)\\

\enddata

\end{deluxetable}

\section{Images}\label{Images}

We provide all of our images to the community online through {\em IRSA}~\footnote{Available http://irsa.ipac.caltech.edu}. This includes our r$^{\prime}$ LFC image and WIRC J, H, and K$_{s}$ images in instrumental units, as well as our IRAC 3.6, 4.5, 5.8, and 8.0~$\mu$m and our MIPS 24, 70, and 160~$\mu$m images in units of MJy~sr$^{-1}$. The 11 intensity images have been fully reduced and calibrated. In addition, we provide coverage and uncertainty maps for the {\it Spitzer} data. The coverage maps are in units of number of BCDs and the uncertainty maps, which can be used to understand the errors, are computed from MOPEX accounting for Poissonian noise as well as a Gaussian component from the read out. For the optical and Near-IR images we provide bad pixel masks, exposure maps, weight and RMS maps, where available. The weight maps are computed in {\it Swarp} and are maps of the local sky background flux variance across the mosaic. RMS maps are computed from the standard deviation of the image multiplied by the square root of normalized coverage map. All images discussed below are in FITS format, unless otherwise stated.

{\em LFC r$^{\prime}$} - The intensity and RMS maps have a size of 455~Mbytes and a pixel size of 0.36$^{\prime\prime}$. We also include a bad pixel map and an exposure map, these last two are in IRAF pl format with sizes of 216~Kbytes and 2.6~Mbytes, respectively.

{\em WIRC} - We provide intensity, weight, and RMS maps. All images have a pixel size of 0.249$^{\prime\prime}$. For J, the image sizes are 360~Mbytes and for H and K$_{s}$ the image sizes are 347~Mbytes.

{\em IRAC} - We provide the intensity maps as well as maps of the coverage and uncertainty produced by MOPEX. The images are 27 or 115~Mbytes and have a pixel sizes of 1.27 or 0.61$^{\prime\prime}$.

{\em MIPS} - As for IRAC, we provide intensity maps, coverage maps and uncertainty maps. For MIPS 24$\mu$m the image sizes are 20~Mbytes and the pixel size is 1.27$^{\prime\prime}$. For MIPS 70$\mu$m the image sizes are 2.4~Mbytes and the pixel size is 4.00$^{\prime\prime}$, and for MIPS 160$\mu$m the image sizes are 515~Kbytes and the pixel size is 8.00$^{\prime\prime}$.

\section{Summary}

In this series of papers we investigate the properties of galaxies in the A1763~-~A1770 supercluster at $z \simeq 0.2$ by using our {\em Spitzer} IR data from 3.6 to 160 $\mu$m, complemented with ground-based spectroscopic and photometric observations. Here we present the photometric data including optical magnitudes from the SDSS and LFC r$^{\prime}$ image, as well as IR magnitudes from WIRC J, H, K$_{s}$, and fluxes from {\em Spitzer} IRAC 3.6, 4.5, 5.8, and 8.0$\mu$m and MIPS 24, 70, and 160$\mu$m. We present a catalog of sources matched to the MIPS 24$\mu$m galaxies and a separate catalog of sources matched to additional MIPS 70$\mu$m galaxies that are not in the MIPS 24$\mu$m FOV. We offer these images and catalogs to the public through {\em IRSA}. Future papers of this series will present spectroscopic follow-up observations, radio and UV imaging of the same field and the analysis of star-formation rates and stellar masses of the infrared sources.

\acknowledgments

This publication makes use of data from observations made with {\em Spitzer} and from the Sloan Digital Sky Survey as well as the Two Micron All Sky Survey. Support for this work was provided by NASA through an award issued by JPL/Caltech. {\em Spitzer} is a space telescope operated by the Jet Propulsion Laboratory, California Institute of Technology, under a contract with NASA.   Funding for the SDSS and SDSS-II has been provided by the Alfred P. Sloan Foundation, the Participating Institutions, the National Science Foundation, the US Department of Energy, NASA, the Japanese Monbukagakusho, the Max Planck Society, and the Higher Education Funding Council of England. The SDSS is managed by the Astrophysical Research Consortium for the Participating Institutions (see list at http://www.sdss.org/collaboration/credits.html). 2MASS is a joint project of the University of Massachusetts and the Infrared Processing and Analysis Center/California Institute of Technology, funded by NASA and the National Science Foundation.

We thank T. Jarrett for his help with the planning of the Near-IR WIRC observations and data reduction. 
We are also grateful to D. Frayer for his help in the reduction
of 70$\mu$m and 160$\mu$m {\em Spitzer} data. We also thank the anonymous referee for thoughtful comments and careful consideration which have lead to an improved manuscript.

{\it Facilities:} Spitzer (MIPS), Spitzer (IRAC), Palomar 200in (WIRC), Palomar 200in (LFC).

\bibliography{a1763}

\end{document}